\documentclass[12pt]{article}
\usepackage{amssymb}
\usepackage{epsf}
\usepackage{epsfig}
\usepackage{cite}

\usepackage{amsmath}

\newcommand{\lsim}{
\mathrel{\hbox{\rlap{\hbox{\lower4pt\hbox{$\sim$}}}\hbox{$<$}}}}

\newcommand{\gsim}{
\mathrel{\hbox{\rlap{\hbox{\lower4pt\hbox{$\sim$}}}\hbox{$>$}}}}

\def\D0{D\O }

\begin{document}
\begin{titlepage}
\vspace*{-0.0truecm}

\begin{flushright}
Nikhef-2010-033
\end{flushright}

\vspace*{0.8truecm}

\begin{center}
\boldmath
{\Large{\bf Extracting $\gamma$ and Penguin Topologies through \\
CP Violation in $B_s^0\to J/\psi K_{\rm S}$

\vspace*{0.3truecm}

}}
\unboldmath
\end{center}

\vspace{0.9truecm}

\begin{center}
{\bf Kristof De Bruyn, Robert Fleischer  \,and\, Patrick Koppenburg}

\vspace{0.5truecm}

{\sl Nikhef, Science Park 105, NL-1098 XG Amsterdam, The Netherlands}
 
\end{center}

\vspace{1.6cm}
\begin{abstract}
\vspace{0.2cm}\noindent
The $B_s^0\to J/\psi K_{\rm S}$ decay has recently been observed by the CDF 
collaboration and will be of interest for the LHCb experiment. This channel will 
offer a new tool to extract the angle $\gamma$ of the unitarity triangle and to control 
doubly Cabibbo-suppressed penguin corrections to the determination of $\sin2\beta$ from the 
well-known $B_d^0\to J/\psi K_{\rm S}$ mode with the help of the $U$-spin symmetry 
of strong interactions. While any competitive determination of $\gamma$ is interesting, 
the latter aspect is particularly relevant as LHCb will enter a territory of precision which 
makes the control of doubly Cabibbo-suppressed Standard-Model corrections mandatory. 
Using the data from CDF and the $e^+e^-$ $B$ factories as a guideline, we explore the 
sensitivity for $\gamma$ and the penguin parameters and point out that the 
$B_s^0$--$\bar B_s^0$ mixing phase $\phi_s$, which is only about $-2^\circ$ in the 
Standard Model but may be enhanced through new physics, is a key parameter 
for these analyses. We find that the mixing-induced CP violation $S(B_s^0\to J/\psi K_{\rm S})$ 
shows an interesting correlation with $\sin\phi_s$, which serves as a target region for
the first measurement of this observable at LHCb.
\end{abstract}

\vspace*{0.5truecm}
\vfill
\noindent
September  2010
\vspace*{0.5truecm}

\end{titlepage}

\thispagestyle{empty}
\vbox{}
\newpage

\setcounter{page}{1}

\section{Introduction}\label{sec:intro}
%
%
As the LHCb experiment at CERN's Large Hadron Collider (LHC) has just started 
to take first physics data, we are entering a new territory in precision flavour physics. 
A particularly exciting feature of the LHCb research programme is the exploration 
of decays of $B_s$ mesons. Among the various $B_s$ decays with a promising physics 
potential is the $B_s^0\to J/\psi K_{\rm S}$ mode, which has recently been observed
by the CDF collaboration at the Tevatron \cite{CDF-obs}. 

\begin{figure}[htbp] 
   \centering
   \includegraphics[width=6.5truecm]{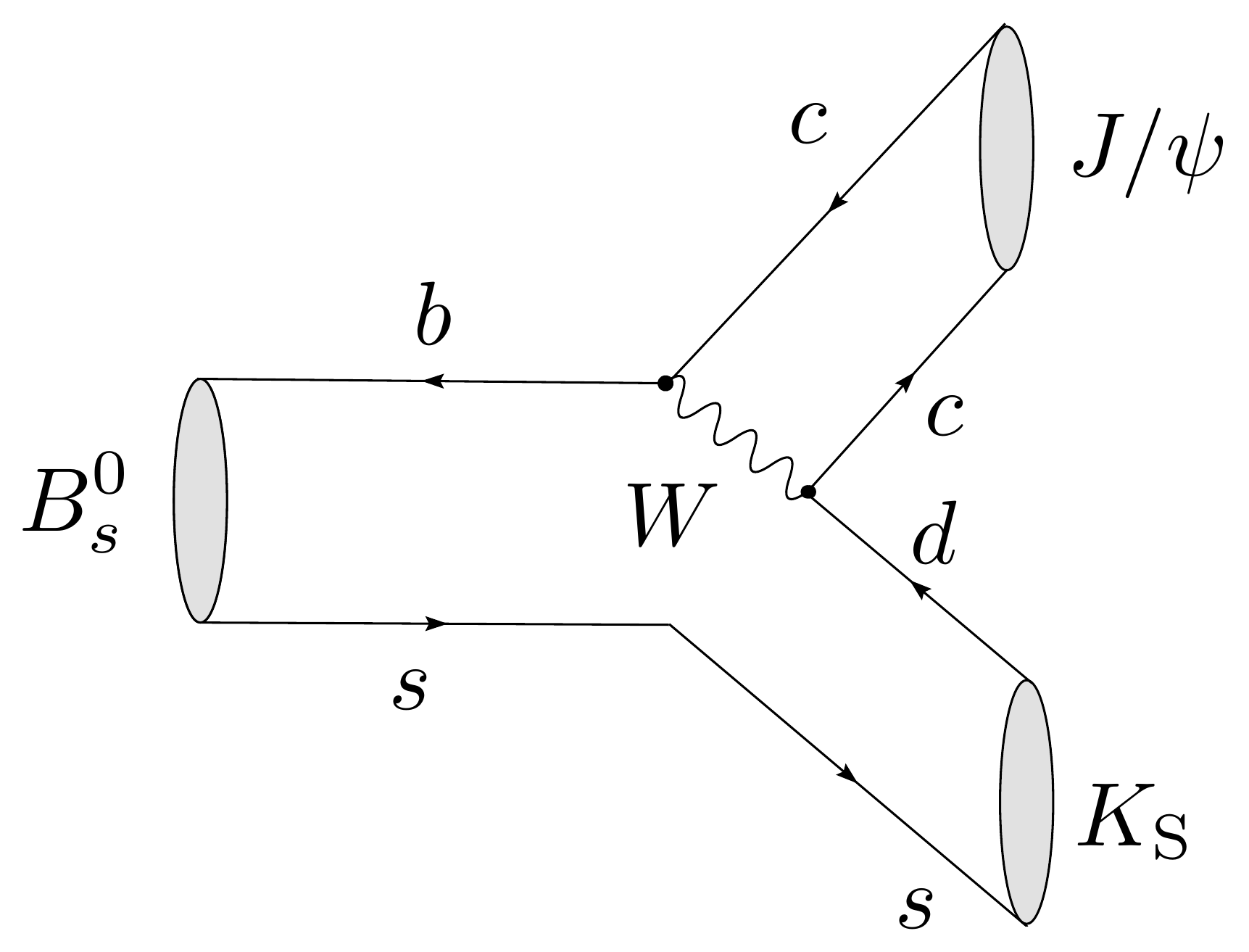}
   \hspace*{1.0truecm} 
   \includegraphics[width=6.5truecm]{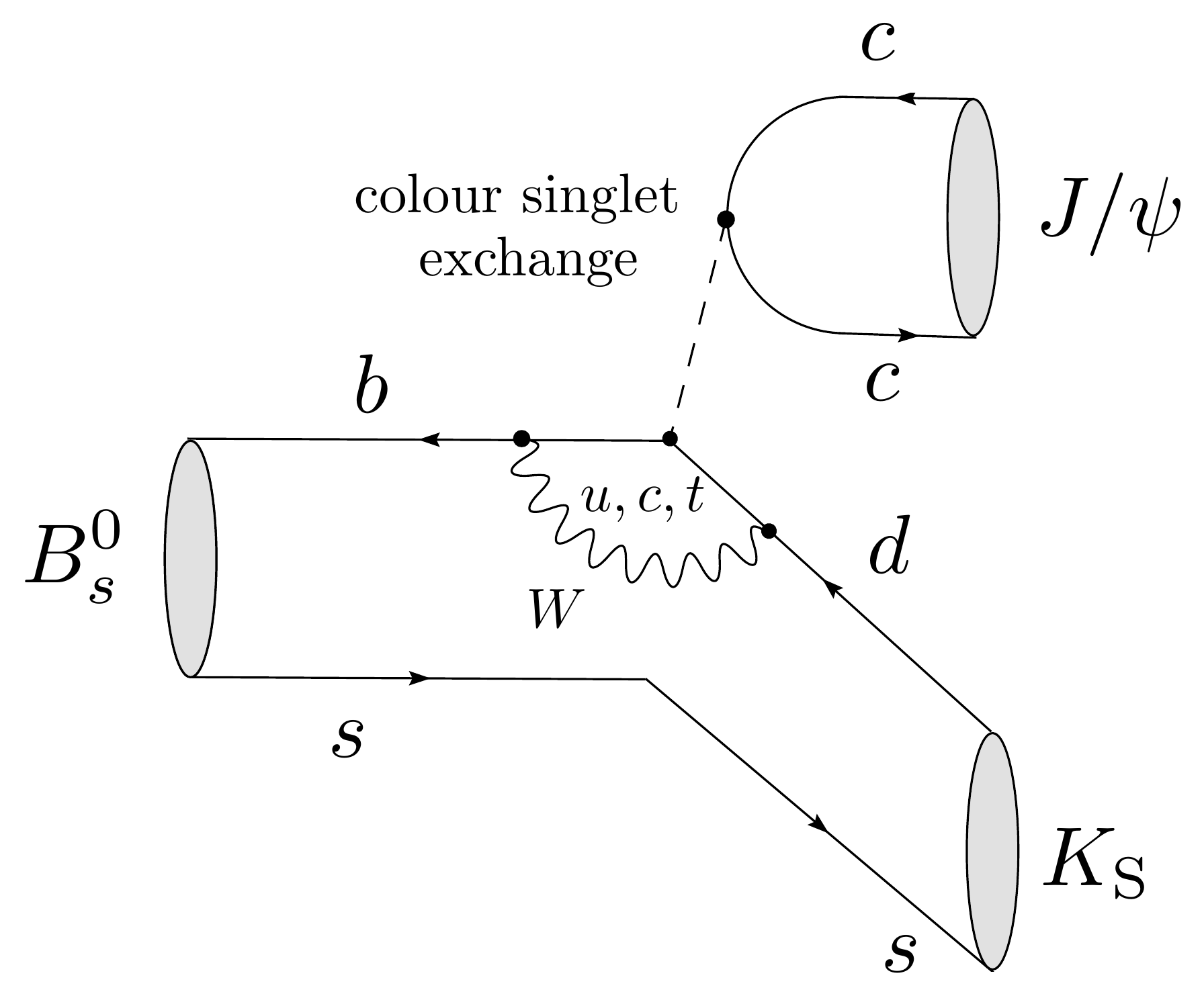} 
     \caption{Decay topologies contributing to the $B_s^0\to J/\psi K_{\rm S}$ channel:
      ``tree" (left)  and ``penguin" (right) diagrams.}\label{fig:BstoJpsiKS}
\end{figure}

In Ref.~\cite{RF-BspsiK}, it was pointed out that the CP-violating asymmetries of this 
channel offer a probe for extracting the angle $\gamma$ of the unitarity triangle (UT). 
Moreover, it was noted that also penguin topologies can be determined, which limit the 
theoretical precision of the measurement of $\sin2\beta$, where $\beta$ is another UT 
angle, by means of the mixing-induced CP violation in the ``golden" decay 
$B_d^0\to J/\psi K_{\rm S}$. In this strategy, the $U$-spin flavour symmetry of 
strong interactions is used, which is a subgroup of $SU(3)_{\rm F}$ relating 
down and strange quarks to each other, and allows us to relate the hadronic 
parameters of the $B_s^0\to J/\psi K_{\rm S}$ and $B_d^0\to J/\psi K_{\rm S}$ 
modes to one another. In Fig.~\ref{fig:BstoJpsiKS}, we show the decay topologies 
of the $B_s^0\to J/\psi K_{\rm S}$ channel; interchanging all down and strange 
quarks, we obtain the $B_d^0\to J/\psi K_{\rm S}$ topologies.

In view of the start of LHCb, we would like to have a closer look at the prospects for
analysing the $B_s^0\to J/\psi K_{\rm S}$ channel at this experiment. Here we will
give a special emphasis to the extraction of the hadronic penguin parameters and
the control of the corresponding uncertainty in the measurement of $\sin2\beta$
from $B_d^0\to J/\psi K_{\rm S}$. In Ref.~\cite{FJFM}, using data on the $B_d^0\to J/\psi\pi^0$
channel (see also \cite{CPS}), 
it was pointed out that such effects tend to lower a tension in the fit of the 
UT between measurements of its side $R_b\propto |V_{ub}/V_{cb}|$ and $\sin2\beta$. 
Following these results, we will explore the prospects for such an analysis by means
of the $B_s^0\to J/\psi K_{\rm S}$ channel as an analysis of $B_d^0\to J/\psi\pi^0$
is considered to be more challenging at LHCb.

Due to the low branching fraction of the $B_s^0\to J/\psi K_{\rm S}$ mode, 
LHCb might be able to get only very crude first information on the corresponding CP
asymmetries by the end of 2011, even though a signal should be clearly visible by then. 
In order to explore the full potential of this decay, 
we will therefore assume two milestones for our feasibility study:
\begin{itemize}
\item an integrated luminosity of $6\:\rm fb^{-1}$, which the LHCb experiment could 
  realistically collect by the end of the second data taking period  of 2014--15 
  (with the LHC having reached the design centre-of-mass energy $\sqrt{s}=14\:\rm TeV$);
\item an integrated luminosity of $100\:\rm fb^{-1}$, which is the current target
  for a subsequent LHCb upgrade.
\end{itemize}

For the measurement  of $\sin2\beta$ through $B_d^0\to J/\psi K_{\rm S}$, 
$(\sin 2\beta)_{J/\psi K_{\rm S}}$, LHCb expects precisions of $0.022$ with 
$2\:\rm fb^{-1}$~\cite{LHCb}, and $0.003$--$0.010$ for $100\:\rm fb^{-1}$, 
depending on to be determined systematic errors ~\cite{LHCbUpgrade}. 
In view
of these impressive numbers, we have to take penguin effects into account in order to match 
the experimental with the theoretical precision. Such an analysis may eventually also 
allow us to pin down new-physics (NP) effects in $B_d^0$--$\bar B_d^0$ mixing; 
the current data still leave space for NP contributions as large as about 50\% of the 
SM contribution, with a CP-violating NP phase below $10^\circ$ 
(see, for instance, Ref.~\cite{BF-mix}).

The outline of this paper is as follows: in Section~\ref{sec:strat}, we discuss the strategy
to extract $\gamma$ and the penguin parameters from the $B_{d,s}\to J/\psi K_{\rm S}$
modes. In Section~\ref{sec:obs}, we have a look at the picture for the relevant observables 
emerging from the current data, and, after introducing the framework for the LHCb feasibility
study in Section~\ref{sec:scen_LHCb}, discuss then in Section~\ref{sec:gam} the prospects for 
the resulting determination of $\gamma$. In Section~\ref{sec:pen}, we focus on the
extraction of the penguin parameters,  and discuss then the control of their impact on the 
measurement of $(\sin 2\beta)_{J/\psi K_{\rm S}}$ in Section~\ref{sec:pen-cont}.
Finally, we summarise our conclusions  and give a brief outlook in Section~\ref{sec:concl}.

\section{The Basic Strategy}\label{sec:strat}
%
%
The $B_s^0\to J/\psi K_{\rm S}$ transition originates from a $\bar b\to\bar c c \bar d$ quark-level
process and has a decay amplitude of the following structure:
\begin{equation}\label{Bs-ampl1}
A(B_s^0\to J/\psi\, K_{\rm S})=\lambda_c^{(d)}\left[A_{\rm T}^{(c)}+
A_{\rm P}^{(c)}\right]+\lambda_u^{(d)}A_{\rm P}^{(u)}
+\lambda_t^{(d)}A_{\rm P}^{(t)}\,,
\end{equation}
where $A_{{\rm T}}^{(c)}$ and $A_{{\rm P}}^{(j)}$ denote strong amplitudes that are related 
to the ``tree" and ``penguin" topologies (with internal $j\in\{u,c,t\}$ quarks), respectively, which
are illustrated in Fig.~\ref{fig:BstoJpsiKS}. On the other hand, the 
\begin{equation}
\lambda_q^{(d)}\equiv V_{qd}^{\phantom{*}}V_{qb}^\ast
\end{equation}
factors contain the relevant combinations of elements of the Cabibbo--Kobayashi--Maskawa
(CKM) matrix. If we make use of the unitarity of the CKM matrix and apply the
Wolfenstein parametrisation \cite{wolf} as generalised in Ref.~\cite{blo}, 
we can write (\ref{Bs-ampl1}) as
\begin{equation}\label{Bs-ampl}
A(B_s^0\to J/\psi\, K_{\rm S})=-\lambda\,{\cal A}\left[1-a e^{i\theta}
e^{i\gamma}\right].
\end{equation}
Here
\begin{equation}
{\cal A}\equiv \lambda^2 A \left[A_{{\rm T}}^{(c)}+A_{{\rm P}}^{(c)}-
A_{{\rm P}}^{(t)}\right]
\end{equation}
and
\begin{equation}\label{a-def}
a e^{i\theta}\equiv R_b
\left[\frac{A_{{\rm P}}^{(u)}-A_{{\rm P}}^{(t)}}{A_{{\rm T}}^{(c)}+
A_{{\rm P}}^{(c)}-A_{{\rm P}}^{(t)}}\right]
\end{equation}
are CP-conserving, ``hadronic" parameters, where 
\begin{equation}
A\equiv \frac{|V_{cb}|}{\lambda^2}=0.809\pm0.026 \quad\mbox{and}\quad
R_b\equiv \left(1-\frac{\lambda^2}{2}\right)\frac{1}{\lambda}\left|\frac{V_{ub}}{V_{cb}}\right|, 
\end{equation}
with $\lambda\equiv|V_{us}|=0.22521\pm0.00083$ are CKM parameters. 
It should be noted that $R_b$ is one of the sides of the UT.

The $B_s^0\to J/\psi K_{\rm S}$ channel is a decay into a CP eigenstate with eigenvalue 
$-1$, and offers a time-dependent CP asymmetry of the following structure:
\begin{eqnarray}\label{acp-def}
a_{\rm CP}(t) \equiv \lefteqn{\frac{\Gamma(B_s^0\to J/\psi K_{\rm S})-
\Gamma(\bar B_s^0\to J/\psi K_{\rm S})}{\Gamma(B_s^0\to J/\psi K_{\rm S})+
\Gamma(\bar B_s^0\to J/\psi K_{\rm S})}=}\nonumber\\
&&\frac{C(B_s^0\to J/\psi K_{\rm S})
\cos(\Delta M_s t)-S(B_s^0\to J/\psi K_{\rm S})\sin(\Delta M_s t)}{\cosh(\Delta\Gamma_st/2)-
{\cal A}_{\rm \Delta\Gamma}(B_s^0\to J/\psi K_{\rm S})\sinh(\Delta\Gamma_st/2)} ,
\end{eqnarray}
where $C(B_s^0\to J/\psi K_{\rm S})$ and $S(B_s^0\to J/\psi K_{\rm S})$ describe direct
and mixing-induced CP violation, respectively, and 
${\cal A}_{\Delta\Gamma}(B_s^0\to J/\psi K_{\rm S})$ is accessible thanks to the
width difference $\Delta\Gamma_s$ of the $B_s$-meson system. In terms of 
CP-violating phases and the hadronic parameters introduced above, these 
observables read as follows:
\begin{eqnarray}
C(B_s^0\to J/\psi K_{\rm S})&=&
\frac{2a\sin\theta\sin\gamma}{1-2a\cos\theta\cos\gamma+a^2}
\label{Adir-def}\\
S(B_s^0\to J/\psi K_{\rm S})&=&
\frac{\sin\phi_s-2a\cos\theta\sin(\phi_s+\gamma)+
a^2\sin(\phi_s+2\gamma)}{1-2a\cos\theta\cos\gamma+a^2}
\label{Amix-def}\\
{\cal A}_{\Delta\Gamma}(B_s^0\to J/\psi K_{\rm S})&=&
\frac{\cos\phi_s-2a\cos\theta\cos(\phi_s+\gamma)+
a^2\cos(\phi_s+2\gamma)}{1-2a\cos\theta\cos\gamma+a^2}\label{AG-def},
\end{eqnarray}
where $\phi_s$ is the CP-violating $B_s^0$--$\bar B_s^0$ mixing phase, which is in
the Standard Model (SM) given by $\phi_s^{\rm SM}=-2\lambda^2\eta=-(2.12\pm0.11)^\circ$
but may well be enhanced by NP effects \cite{BF-mix}.

It should be noted that these quantities are not independent from one another, satisfying
the relation
\begin{equation}
\left[C(B_s^0\to J/\psi K_{\rm S})\right]^2+\left[S(B_s^0\to J/\psi K_{\rm S})\right]^2+
\left[{\cal A}_{\Delta\Gamma}(B_s^0\to J/\psi K_{\rm S})\right]^2=1.
\end{equation}
However, we have still another independent observable at our disposal, which is the
CP-averaged rate \cite{RF-BspsiK}
\begin{equation}\label{R-def}
\langle\Gamma(B_s^0\to J/\psi K_{\rm S})\rangle\equiv\Phi^s_{J/\psi K_{\rm S}}\times|{\cal N}|^2\times
\left[1-2 a\cos\theta\cos\gamma+a^2\right],
\end{equation}
where $\Phi^s_{J/\psi K_{\rm S}}$ denotes a phase-space factor, and 
${\cal N}\equiv-\lambda {\cal A}$.

In the case of the decay $B_d^0\to J/\psi K_{\rm S}$, the counterpart of the
penguin parameter $ae^{i\theta}$ enters with the doubly Cabibbo-suppressed
parameter $\epsilon\equiv\lambda^2/(1-\lambda^2)=0.053$ as follows:
\begin{equation}\label{Bd-ampl2}
A(B_d^0\to J/\psi\, K_{\rm S})=\left(1-\frac{\lambda^2}{2}\right){\cal A'}
\left[1+\epsilon a'e^{i\theta'}e^{i\gamma}
\right],
\end{equation}
where the hadronic parameters are defined in analogy to the $B_s^0\to J/\psi K_{\rm S}$
case \cite{RF-BspsiK}, and the primes remind us that we are dealing with a $\bar b\to \bar s$
mode. The corresponding observables can be obtained from the expressions given above
by simply making the following replacements:
\begin{equation}
a\to-a', \quad \theta\to\theta', \quad {\cal A}\to {\cal A}', \quad
{\cal N}\to {\cal N}'\equiv -\frac{1}{\sqrt{\epsilon}}
\frac{\cal A'}{\cal A}{\cal N}.
\end{equation}
Concerning the mixing-induced CP asymmetry $S(B_d^0\to J/\psi K_{\rm S})$, we have 
to replace, in addition, the $B_s^0$--$\bar B_s^0$ mixing phase by its $B_d$-meson
counterpart $\phi_d$, which is in the SM given by $\phi_d^{\rm SM}=2\beta$.

Thanks to the $\epsilon$ suppression of the penguin parameter in (\ref{Bd-ampl2}), the
$B^0_d\to J/\psi K_{\rm S}$ decay is referred to as the ``golden" channel for the measurement
of $\phi_d$. In the $B$-factory era of the last decade, it was justified in terms of accuracy to
simply neglect this term. On the other hand, we are entering a new territory for precision $B$ 
physics in the LHC era which makes it mandatory to take the penguin effects into account. 
This was already anticipated in 1999 in Ref.~\cite{RF-BspsiK}, where it was proposed to 
use the $B^0_s\to J/\psi K_{\rm S}$ channel to accomplish this task and to
extract the angle $\gamma$ of the UT, exploiting the feature that the decays 
\mbox{$B_s^0\to J/\psi K_{\rm S}$} and $B_d^0\to J/\psi K_{\rm S}$ are related 
to each other through an interchange of all down and strange quarks, i.e.\ by the $U$-spin
flavour symmetry of strong interactions. The relevant observables are the 
CP asymmetries $C(B_s^0\to J/\psi K_{\rm S})$ and $S(B_s^0\to J/\psi K_{\rm S})$ as 
well as the following ratio of the CP-averaged rates:
\begin{equation}\label{H-expr}
H\equiv \frac{1}{\epsilon} 
\frac{\langle\Gamma(B_s^0\to J/\psi K_{\rm S})\rangle}{\langle
\Gamma(B_d^0\to J/\psi K_{\rm S})\rangle}
\left|\frac{{\cal A}'}{{\cal A}}\right|^2
\frac{\Phi^d_{J/\psi K_{\rm S}}}{\Phi^s_{J/\psi K_{\rm S}}}=
\frac{1-2a\cos\theta\cos\gamma+a^2}{1+
2\epsilon a'\cos\theta'\cos\gamma+\epsilon^2 a'^2}.
\end{equation}

The $U$-spin flavour symmetry implies
\begin{equation}\label{rel-1}
a=a', \quad \theta=\theta',
\end{equation}
as well as
\begin{equation}
{\cal A'}= {\cal A}.
\end{equation}
The three observables depend then on $\gamma$ as well as $a$ and
$\theta$. Consequently, we can convert the measured values of $C(B_s^0\to J/\psi K_{\rm S})$, 
$S(B_s^0\to J/\psi K_{\rm S})$ and $H$ into $\gamma$, $a$ and $\theta$. In Sections~\ref{sec:gam}
and \ref{sec:pen}, we discuss in detail how the implementation of this strategy works in 
practise at the LHCb experiment.

\section{Picture from Current Data}\label{sec:obs}
%
%
The CDF collaboration has recently reported the observation of the $B_s^0\to J/\psi K_{\rm S}$
channel, with the following result \cite{CDF-obs}:
\begin{equation}\label{CDF-rat}
\frac{\mbox{BR}(B_s^0\to J/\psi K_{\rm S})}{\mbox{BR}(B_d^0\to J/\psi K_{\rm S})}=
0.0405 \pm 0.0070 \mbox{(stat.)} \pm 0.0041 \mbox{(syst.)} \pm 0.0050 \mbox{(frag.)},
\end{equation}
where the last error is associated with the ratio $f_s/f_d$ of the $B_s$ and $B_d$ fragmentation
functions. Using the value $\mbox{BR}(B_d^0\to J/\psi K^0)=(8.71 \pm 0.32) \times 10^{-3}$ 
of the Particle Data Group (PDG) \cite{PDG} yields
\begin{equation}
\mbox{BR}(B_s^0\to J/\psi \bar K^0)=(3.53 \pm 0.61 \mbox{(stat.)} \pm  0.35 \mbox{(syst.)}
\pm 0.43 \mbox{(frag.)} \pm 0.13 \mbox{(PDG)}) \times 10^{-5}.
\end{equation}

It is interesting to compare this result with the branching ratio of the $B_d^0\to J/\psi \pi^0$
channel. If we neglect penguin annihilation and exchange topologies, which contribute to 
$B_d^0 \to J/\psi \pi^0$ but have no counterpart in $B_s^0 \to J/\psi \bar K^0$  and are 
expected to play a minor role (which can be probed through $B_s^0\to J/\psi \pi^0$), the
$SU(3)$ flavour symmetry applied to the spectator $d$ and $s$ quarks implies 
\begin{equation}\label{Xi-def}
\Xi_{SU(3)}\equiv
\frac{\mbox{BR}(B_s^0\to J/\psi \bar K^0)}{2\mbox{BR}(B_d^0\to J/\psi\pi^0)}
\frac{\tau_{B_d}}{\tau_{B_s}}\frac{\Phi^d_{J/\psi \pi^0}}{\Phi^s_{J/\psi K_{\rm S}}}
\, \stackrel{SU(3)}{\longrightarrow}1, 
\end{equation}
where the factor of 2 is associated with the wave function of the $\pi^0$, whereas the
$\tau_{B_d}$ and $\tau_{B_s}$ denote the $B_d$ and $B_s$ lifetimes, respectively.
Using the PDG values $\tau_{B_d}/\tau_{B_s}=1.036\pm0.018$ and 
$\mbox{BR}(B_d^0\to J/\psi\pi^0)=(1.76\pm0.16)\times10^{-5}$, we obtain
\begin{equation}\label{Xi-res}
\Xi_{SU(3)}=1.01\pm0.25,
\end{equation} 
which agrees very well with the $SU(3)$ expectation in (\ref{Xi-def}), although the error
is still sizable.

In the following analysis, we will use the hadronic parameters extracted from the CP violation
data for the $B_d^0\to J/\psi \pi^0$ channel as a guideline, using the numerical values
obtained in Ref.~\cite{FJFM}. Making the same assumptions as in (\ref{Xi-def}), we have
then
\begin{equation}\label{eq:a_theta_estimate}
a\in[0.15,0.67], \quad \theta\in[174,213]^\circ,
\end{equation} 
which  correspond to $\gamma=(65\pm10)^\circ$. The numerical value in (\ref{Xi-res}) 
gives us confidence in this procedure. The observables of the 
$B_s^0\to J/\psi K_{\rm S}$ channel then read as 
\begin{equation}
H=1.53\pm0.31,
\end{equation}
and
\begin{equation}
C\equiv C(B_s^0\to J/\psi K_{\rm S})=C(B_d^0\to J/\psi\pi^0)= -0.10\pm0.13.
\end{equation}
Concerning the mixing-induced CP violation $S\equiv S(B_s^0\to J/\psi K_{\rm S})$, 
we can calculate this observable with the help of Eq.~(\ref{Amix-def})  as a function of the 
$B_s^0$--$\bar B_s^0$ mixing phase $\phi_s$ for given values of $H$, $C$ 
and $\gamma$. The corresponding formulae are given as follows:
\begin{equation}
2a\cos\theta=\frac{1-H+(1-\epsilon^2H)a^2}{(1+\epsilon H)\cos\gamma}
\end{equation}
\begin{equation}
1-2a\cos\theta\cos\gamma+a^2=\left[\frac{(1+\epsilon)(1+\epsilon a^2)}{1+\epsilon H} \right]H,
\end{equation}
where
\begin{equation}
a^2=P\pm\sqrt{P^2-Q}
\end{equation}
with
\begin{equation}
P=\frac{2\left[(1+\epsilon H)\sin\gamma\cos\gamma\right]^2-(1-H)(1-\epsilon^2H)\sin^2\gamma
-\epsilon\left[(1+\epsilon)H C\cos\gamma\right]^2}{\left[(1-\epsilon^2H)\sin\gamma\right]^2
+\left[\epsilon(1+\epsilon)HC\cos\gamma\right]^2}
\end{equation}
\begin{equation}
Q=\frac{\left[(1+\epsilon)HC\cos\gamma\right]^2+
\left[(1-H)\sin\gamma\right]^2} {\left[(1-\epsilon^2H)\sin\gamma\right]^2
+\left[\epsilon(1+\epsilon)HC\cos\gamma\right]^2}.
\end{equation}
In these expressions, we have used the $U$-spin relation (\ref{rel-1}) but did not make any
other approximation. If we assume the SM, we arrive at the following prediction:
\begin{equation}
S(B_s^0\to J/\psi K_{\rm S})|_{\rm SM}=0.54\:^{+\:0.14}_{-\:0.25}
|_{H}\:^{+\:0.00}_{-\:0.01}|_C\:^{+\:0.16}_{-\:0.13}|_\gamma=0.54\:^{+\:0.21}_{-\:0.28}.
\end{equation}

As we have noted above, NP may well enter the $B_s^0\to J/\psi K_{\rm S}$ channel
through contributions to $B_s^0$--$\bar B_s^0$ mixing, yielding a value
of $\phi_s$ different from $\phi_s^{\rm SM}$. The most recent compilation of
the CDF and \D0 collaborations using measurements of CP violation in $B_s^0\to J/\psi \phi$
\cite{DDFN} can be found in Refs.~\cite{CDF-phis} and \cite{D0-phis}, respectively. 
Unfortunately, the situation is not conclusive, although the CDF and \D0 data are 
consistent with each other:  the CDF collaboration finds the ranges
$\phi_s\in [-59.6^\circ,-2.29^\circ]\sim-30^\circ \, \lor \, [-177.6^\circ,-123.8^\circ]\sim-150^\circ$
(68\% C.L.), while the \D0 collaboration gives a best fit value around $\phi_s\sim -45^\circ$, 
taking also information from the dimuon charge asymmetry and the measured 
$B_s^0\to D_s^{(*)+}D_s^{(*)-}$ branching ratio into account. 

\begin{figure}[htbp] 
   \centering
   \includegraphics[width=3in]{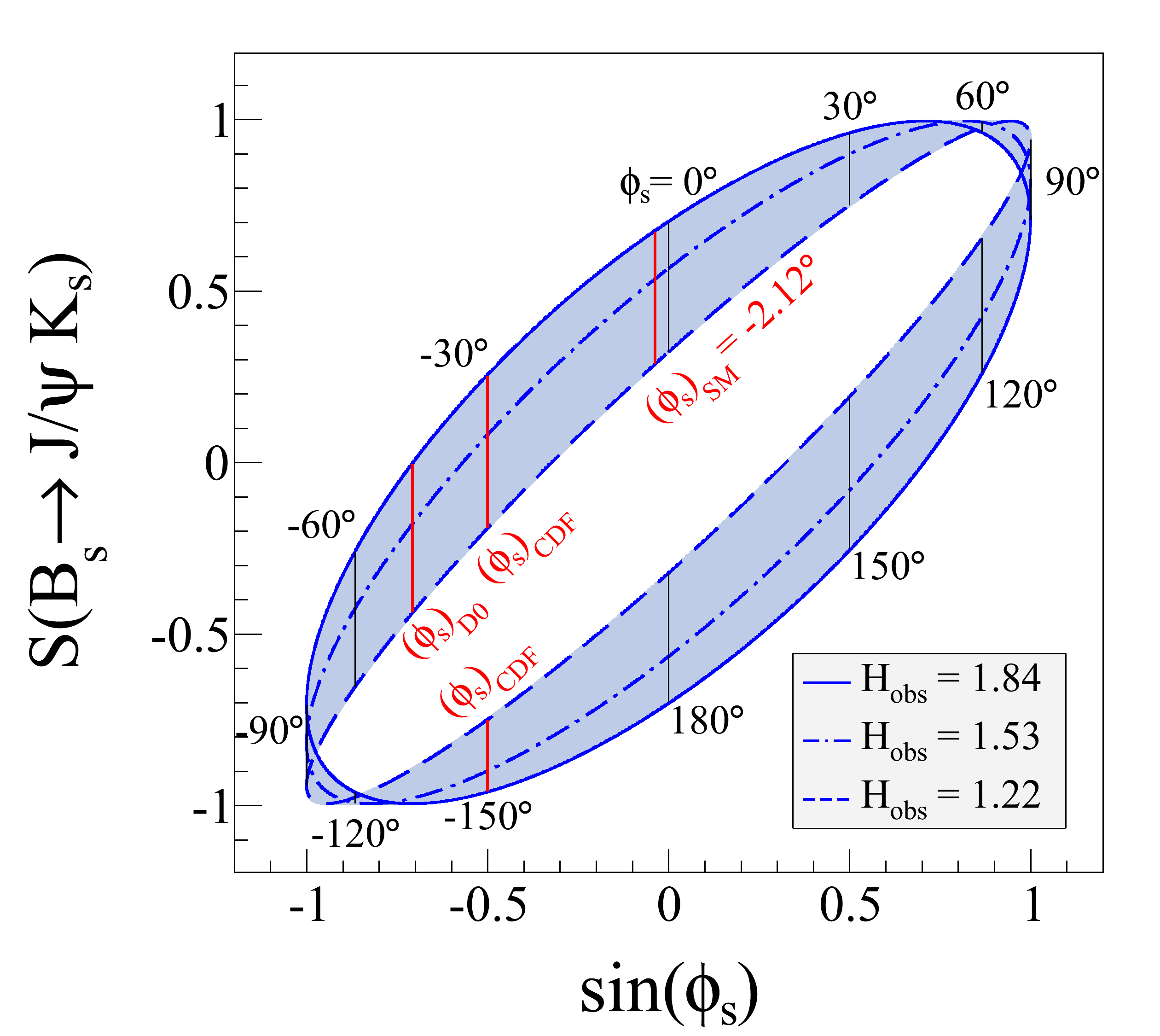}
   \includegraphics[width=3in]{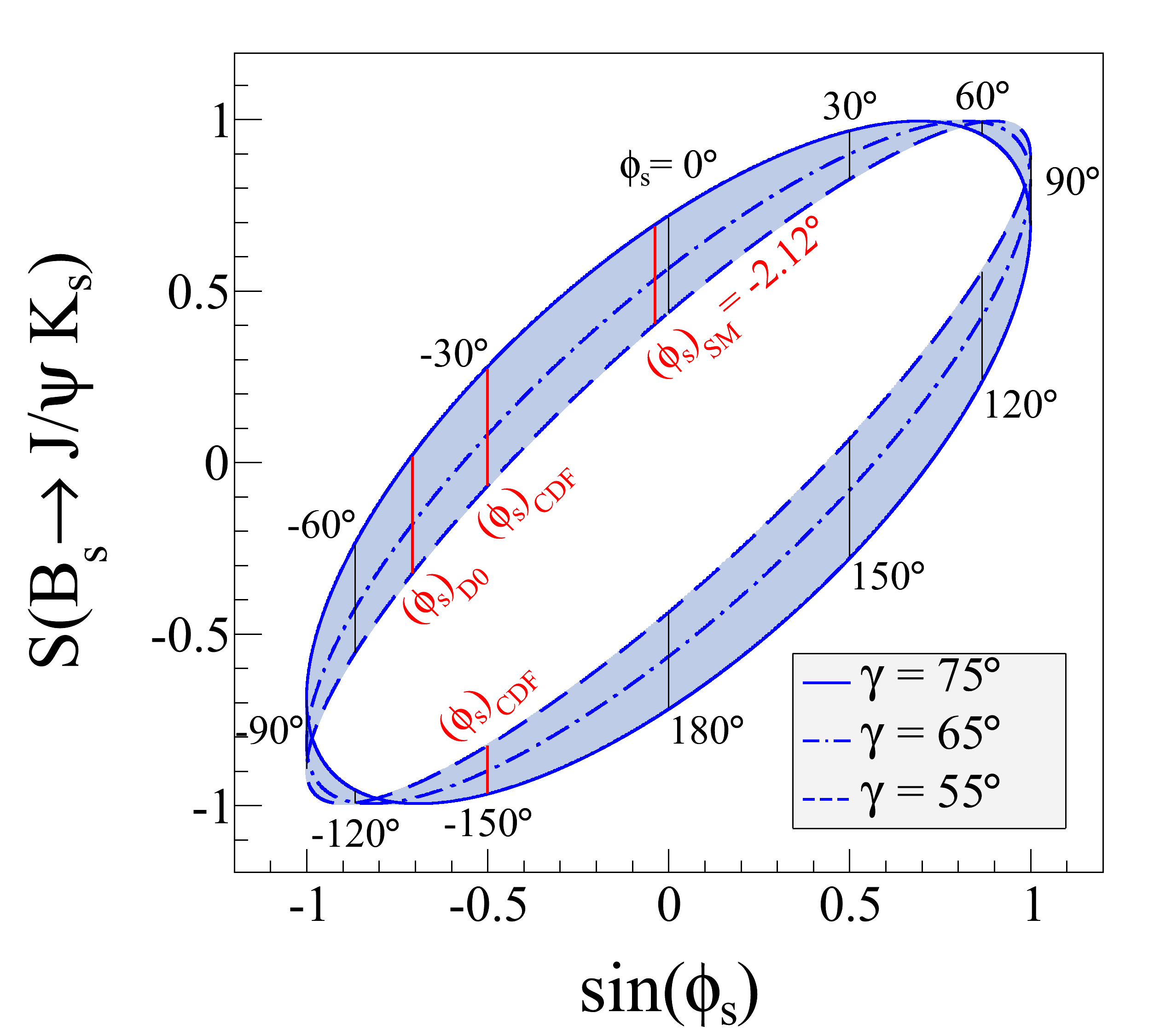}
   \caption{Correlation between $\sin\phi_s$ and $S(B_s^0\to J/\psi K_{\rm S})$ for different values
   of $H$ (left panel, with $\gamma=65^\circ$) and different values of $\gamma$ (right panel, with
   $H=1.53$). The impact of a variation of $C=-0.10\pm0.13$ is small.}
   \label{fig:corr}
\end{figure}

Interestingly, we can also probe this kind of NP through the mixing-induced CP violation
in $B_s^0\to J/\psi K_{\rm S}$. In Fig.~\ref{fig:corr}, we show the correlation
between $S(B_s^0\to J/\psi K_{\rm S})$ and $\sin\phi_s$, which can be determined from
the time-dependent angular analysis of the $B_s^0\to J/\psi\phi$ channel \cite{DDFN}. 
In this figure, we assume that NP is only present in $B^0_s$--$\bar B^0_s$ mixing and
have used the formulae give above to calculate curves for different values
of $H$ and $\gamma$. We observe the following features:
\begin{itemize}
\item Thanks to the penguin contributions, we can distinguish between the twofold ambiguity
for $\phi_s$ originating from the measured value of $\sin\phi_s$, in particular also between
$\phi_s\sim 0^\circ$ (as in the SM) and $\phi_s\sim 180^\circ$.
\item For $\phi_s\sim-30^\circ$ and $\phi_s\sim-45^\circ$, the mixing-induced CP asymmetry is $\sim0$ while it
approaches $-1$ for $\phi_s\sim-150^\circ$. Consequently, 
we could then also clearly identify the scenarios corresponding to the CDF and \D0 data
discussed above. 
\end{itemize}
In view of these observations, already first experimental information on the mixing-induced 
CP violation in $B_s^0\to J/\psi K_{\rm S}$ would be very interesting to complement the 
analyses of CP violation in $B_s^0\to J/\psi \phi$. Needless to note that this feature relies on 
sizable penguin contributions to the $B_s^0\to J/\psi K_{\rm S}$ channel. 

\section{Framework for the LHCb Feasibility Study}\label{sec:scen_LHCb}
%
%
With an anticipated integrated luminosity of $1\:\rm fb^{-1}$ by the end of 2011, the LHCb
experiment expects to be able to measure $\sin\phi_s$ with a precision 
between $0.04$ and $0.07$~\cite{Roadmap}. The $B_s^0\to J/\psi K_{\rm S}$ decay 
should then already be clearly visible with the following uncertainties:
\begin{equation}
\frac{\mbox{BR}(B_s^0\to J/\psi K_{\rm S})}{\mbox{BR}(B_d^0\to J/\psi K_{\rm S})}=
0.0403 \pm 0.0065 \mbox{(stat.)} \pm 0.0029 \mbox{(frag.)},
\end{equation}
based on an estimated event yield of $60\,000$ $B_d^0\to J/\psi K_{\rm S}$ and $700$ $B_s^0\to J/\psi K_{\rm S}$ decays, and
where the error on the fragmentation fractions is based on the method proposed in 
Ref.~\cite{fd_fs}. Concerning the CP-violating observables, we expect a sensitivity of
$\Delta C = \Delta S = 0.34$, which might allow us to get first information on these
quantities.

In order to explore the feasibility of the $B_s^0\to J/\psi K_{\rm S}$ measurement at LHCb with higher integrated luminosities, we perform a toy Monte Carlo (MC) analysis. For this study we assume the following situation: Already with an integrated luminosity of $1\:\rm fb^{-1}$ we can pin down the $B_s^0$--$\bar B_s^0$ mixing phase. Hence, for the two milestone scenarios it will be well known. In the following discussion we will consider the SM case $\phi_s = -2.12^{\circ}$. Secondly, the used signal and background yields are extrapolated from Ref.~\cite{LHCb}.  The number of $B_d^0\rightarrow J/\psi K_{\rm S}$ events listed there is scaled, using the measured $B_s^0\to J/\psi K_{\rm S}$ branching fraction, to give an expected yield of $4000$ and $70\,000$ $B_s^0\rightarrow J/\psi K_{\rm S}$ events for the integrated luminosities of $6\:\rm fb^{-1}$ and  $100\:\rm fb^{-1}$, respectively. 
These event yields are based on an estimated inclusive $J/\psi$ cross section of $\sigma_{\text{Incl.} J/\psi} = 262\:\mu\mathrm{b}$ at $\sqrt{s} = 14\: \rm TeV$, the LEP fragmentation functions published by HFAG \cite{HFAG}, and a $b\bar{b}$ cross section at $\sqrt{s} =14\: \rm TeV$ of $\sigma_{b\bar{b}} = 568\:\mu\mathrm{b}$, which was extrapolated from the recently measured $b\bar{b}$ cross section of $\sigma_{b\bar{b}} = (284\pm20\:(\mathrm{stat.})\pm 49\:(\mathrm{syst.}))\:\mu\mathrm{b}$ at $\sqrt{s} = 7\: \rm TeV$ \cite{bbcross}.

\begin{table}[h]
\center
\begin{tabular}{|p{4cm}|p{2.3cm}||p{4cm}|p{2.3cm}|}
\hline
Parameter & Value  & Parameter & Value \\
\hline
$\sigma_{b\bar{b}}$ & $568\:\mu\mathrm{b}$ & $B_s$ lifetime resolution & 0.041 ps\\
$\sigma_{\text{Incl.} J/\psi}$ & $262\:\mu\mathrm{b}$ & mistag fraction & 0.36\\
$f_d/f_s$ & $3.55\pm 0.26$ & $B$ mass resolution & 15.2 MeV/c$^2$\\
\hline
\end{tabular}
\caption{Overview of the parameters used by the toy MC.}
\label{tab:MCpar}
\end{table}

For the MC analysis the reconstructed mass and lifetime distribution of both signal and background are modelled.
The former is used to obtain the errors on the CP observables $C$ and $S$, while the latter is used for the determination of the statistical error on $H$. 
The error on $H$ coming from the ratio $f_s/f_d$ of fragmentation functions, affecting $H$ via \eqref{CDF-rat}, is again determined using the strategy proposed in Ref.~\cite{fd_fs}. The most import parameters used  by the toy MC to describe both distributions are listed in Table \ref{tab:MCpar}.

\begin{figure}[htbp] 
   \centering
   \includegraphics[width=3in]{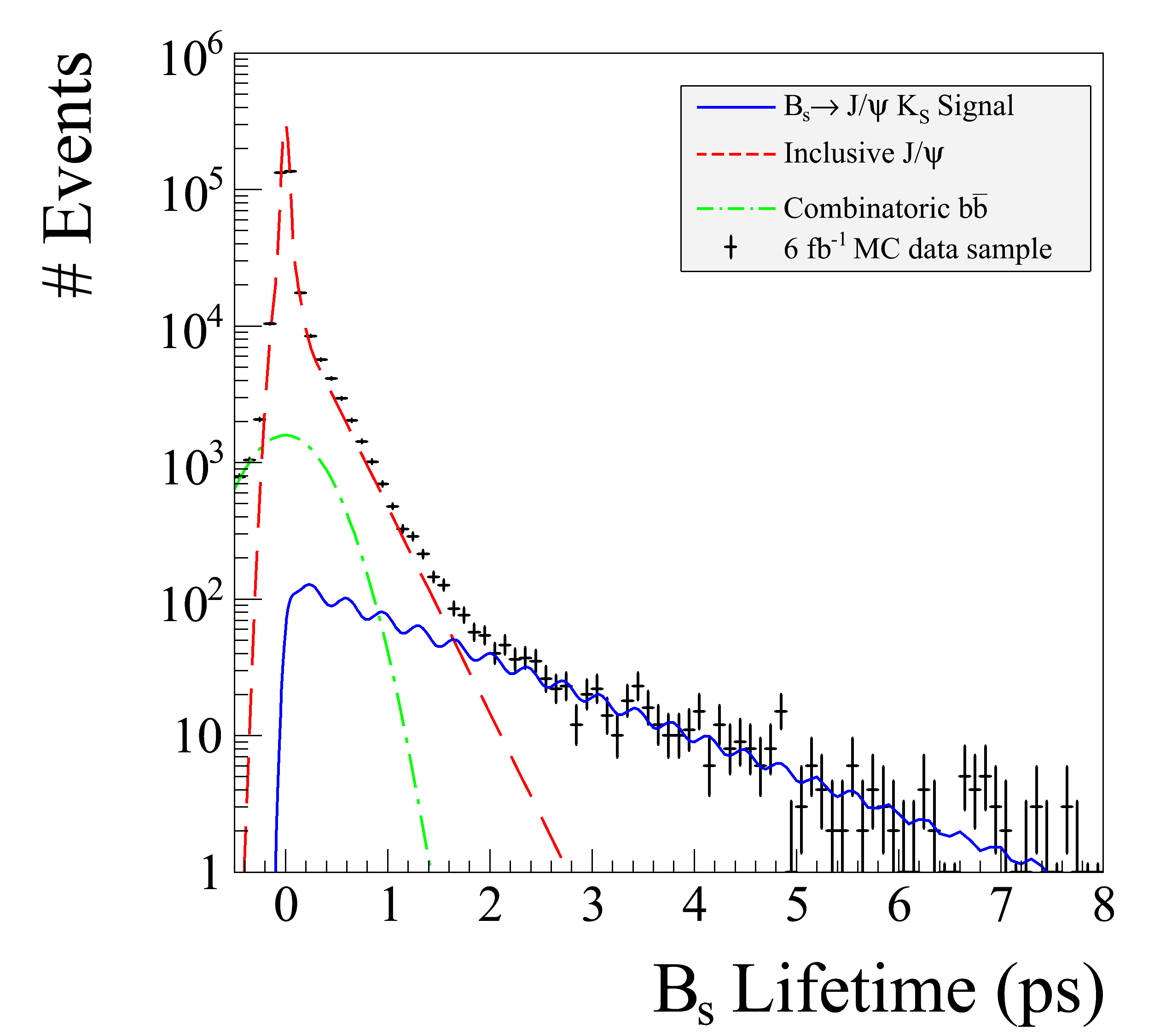}
   \includegraphics[width=3in]{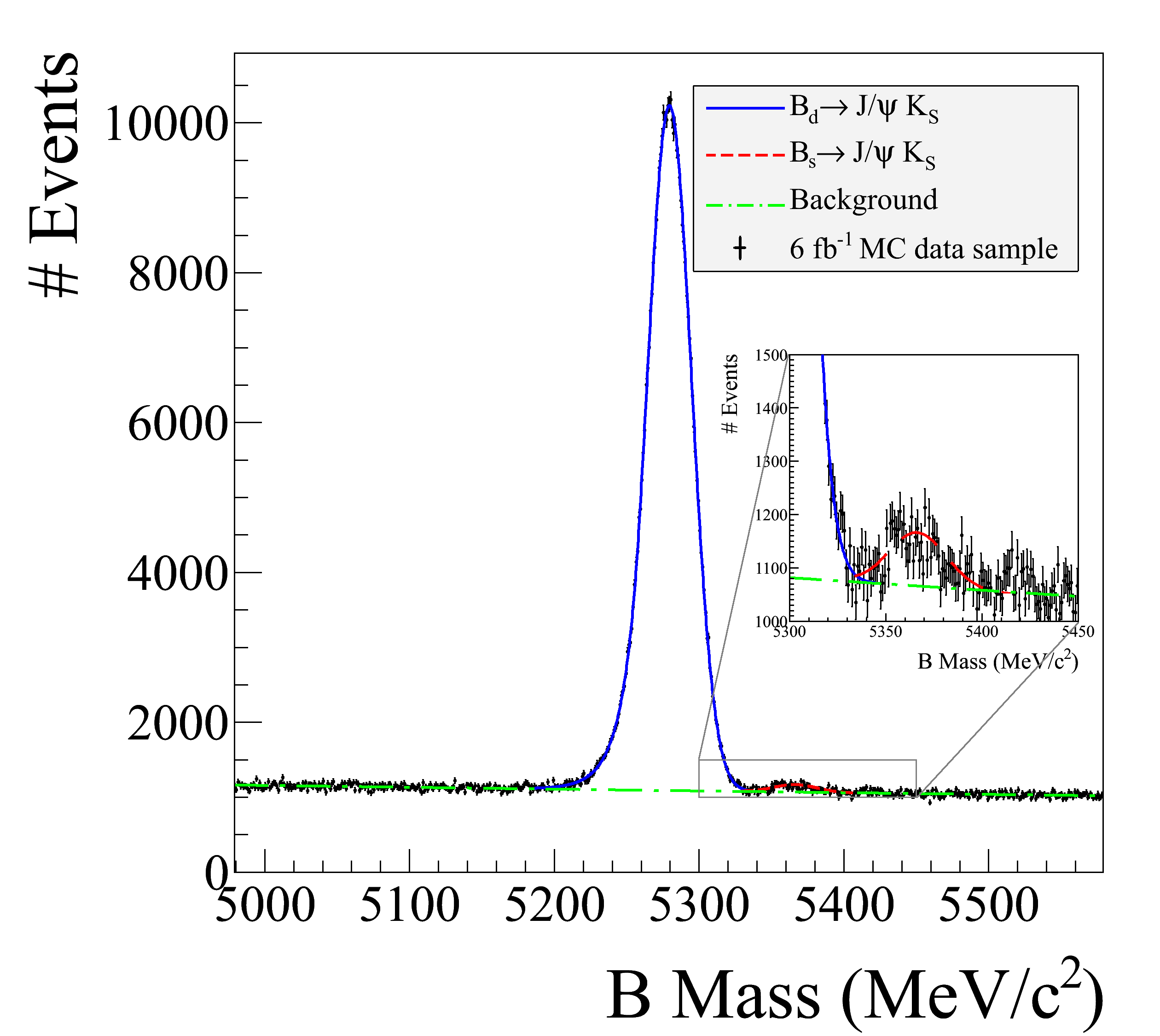}\\
   \includegraphics[width=3in]{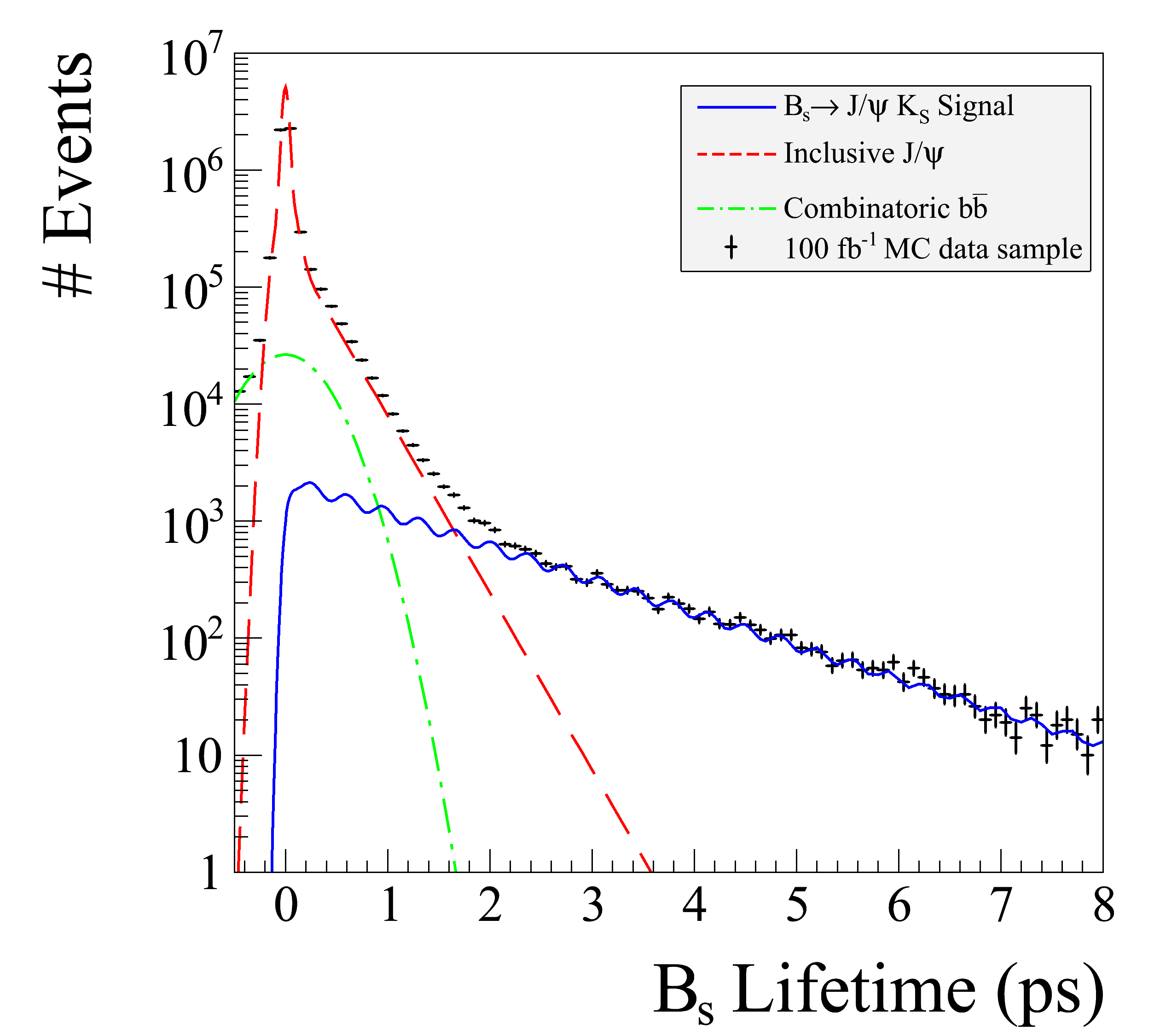}
   \includegraphics[width=3in]{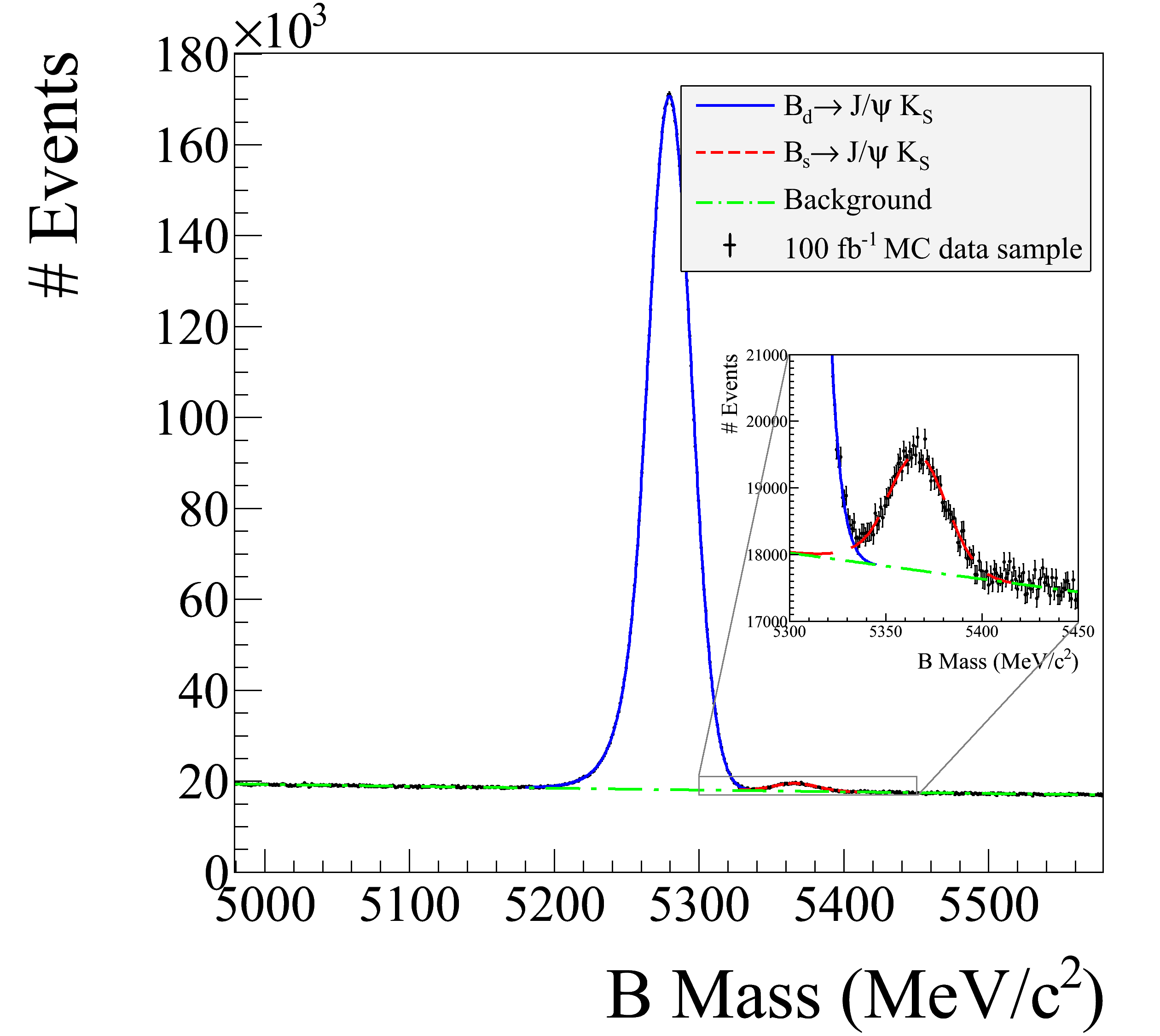}
   \caption{The lifetime distribution for an originally tagged $B_s$ meson (left) and the total mass distribution  (right) for the decay $B_s^0\rightarrow J/\psi K_{\rm S}$ and its background, as used in the MC analysis. This shows good sensitivity to the signal in the long lifetime region. The number of events correspond to a data sample of $6\:\rm fb^{-1}$ (top) and $100\:\rm fb^{-1}$ (bottom) respectively.}
   \label{fig:MC_model}
\end{figure}

The modelled $B_s$ lifetime distribution, given in the left panels of Fig.~\ref{fig:MC_model}, has three major components: the $B_s^0\rightarrow J/\psi K_{\rm S}$ signal (plotted for $B_s$ and $\bar{B}_s$ combined), an inclusive $J/\psi$ background and a combinatoric $b\bar{b}$ background. The $B_s$ and $\bar{B}_s$ signals are described individually using the time evolution equations (11) and (12) in Ref.~\cite{RF-BspsiK}, respectively, and include a dilution factor due to mistagging of the original $B$ meson. The inclusive $J/\psi$ sample contains both prompt and non-prompt (e.g.\ from $B$ decays) $J/\psi$ candidates, which results in an asymmetric distribution. The combinatoric $b\bar{b}$ background describes all other background from $b\bar{b}$ events; other sources of background are considered negligible. All lifetime distributions are convoluted with a Gaussian resolution model, which causes some events to have a negative lifetime. The resolution for the $B_s^0\to J/\psi K_{\rm S}$ signal is $\sigma = 0.041\:\rm ps$ \cite{LHCb}.

The modelled mass distribution, also given in Fig.~\ref{fig:MC_model}, describes the two $B$ mass peaks ($B_d$ on the left, and $B_s$ on the right and in the insert) and an exponential background for both the inclusive $J/\psi$ and combinatoric $b\bar{b}$ sample.

\begin{figure}[htbp] 
   \centering
   \includegraphics[height=3in]{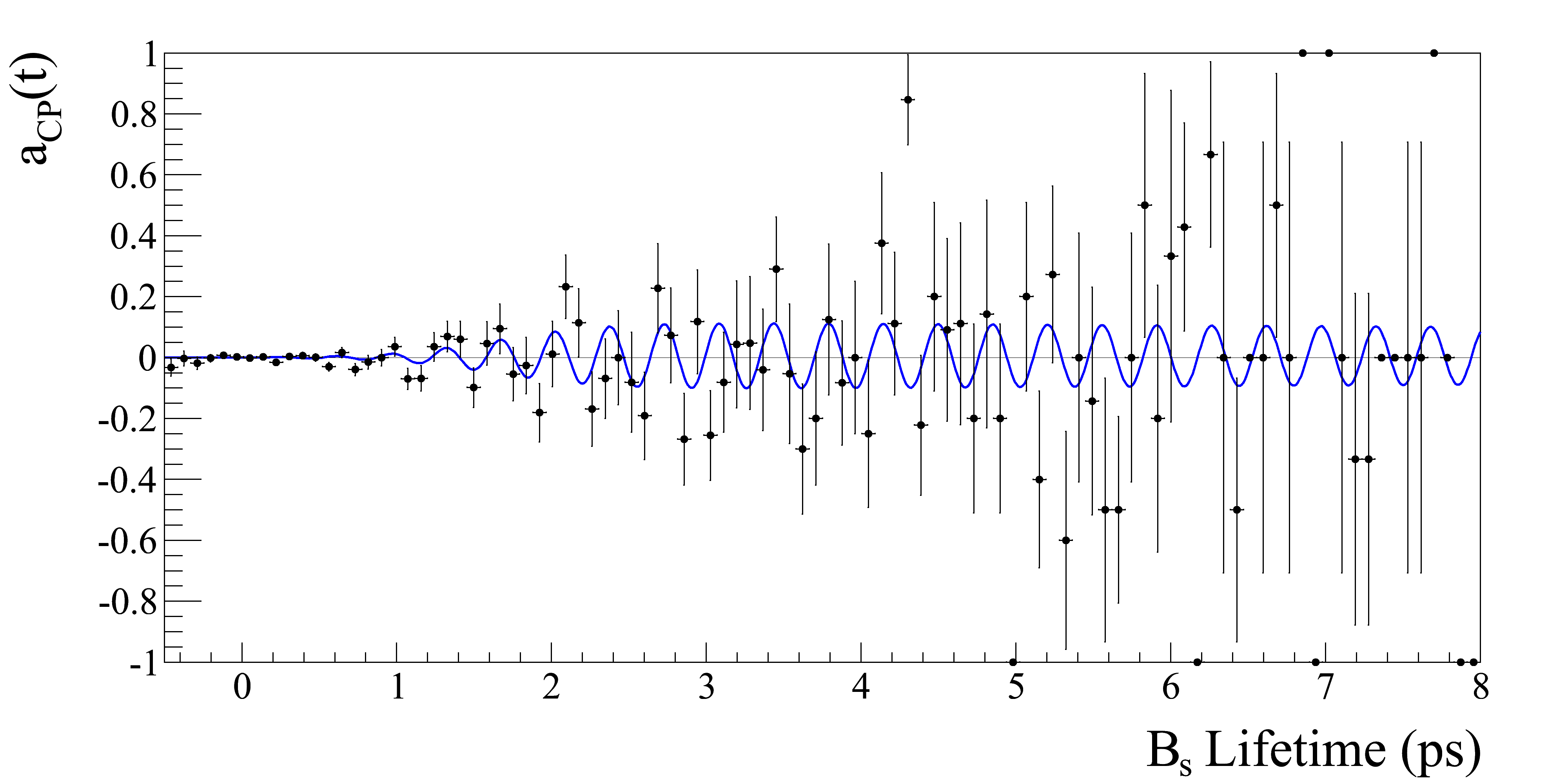}\\
   \includegraphics[height=3in]{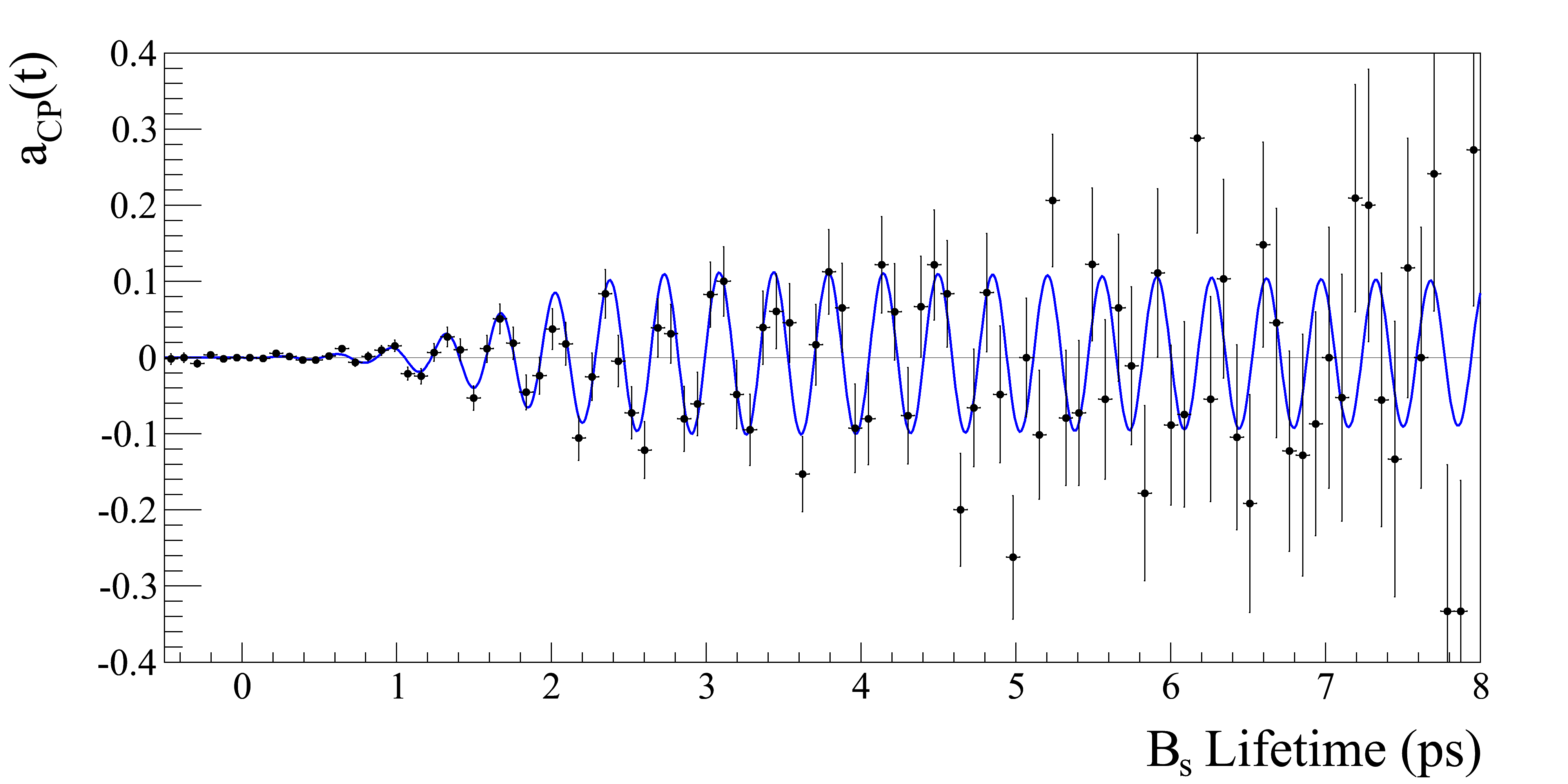}
   \caption{Toy MC simulation of the time-dependent CP asymmetry $a_{\rm CP}(t)$ defined in equation \eqref{acp-def} for the $6\:\rm fb^{-1}$ (top) and $100\:\rm fb^{-1}$ (bottom) scenarios.}
   \label{fig:CPasym}
\end{figure}

The signal yield is small compared to the number of background events, as can be seen from the inserts in Fig.~\ref{fig:MC_model},
making it hard to distinguish the $B_s$ mass peak from the background. This is the main reason for choosing the high luminosity 
scenarios of $6\:\rm fb^{-1}$ and $100\:\rm fb^{-1}$. Nonetheless, the lifetime distribution shows that the background is 
predominantly prompt. The long lifetime region ($t>2$ ps) is therefore sufficiently sensitive to 
$B_s^0\rightarrow J/\psi K_{\rm S}$ in order to extract $C$ and $S$.
This is confirmed by the plot of the time-dependent CP asymmetry $a_{\rm CP}(t)$ defined in equation \eqref{acp-def} and given in Fig.~\ref{fig:CPasym} for our two milestone scenarios.

The toy MC study is mainly used to estimate the errors on the relevant observables. The actual values of these observables are determined by their dependency on $a$, $\theta$ and $\gamma$, given in equations \eqref{Adir-def}, \eqref{Amix-def} and \eqref{H-expr}. The input values for $a$, $\theta$ and $\gamma$ were taken from the $B_d^0\rightarrow J/\psi\pi^0$ analysis whose result was already given in \eqref{eq:a_theta_estimate}. We take the central values of 
these intervals,
\begin{equation}\label{eq:MC_input}
a=0.41\quad,\quad\theta = 194^{\circ}\quad,\quad\gamma = 65^{\circ},
\end{equation}
which result  in the following picture for the statistical uncertainties at LHCb:
\begin{eqnarray}
6~\text{fb}^{-1}: & 
\Delta H= 0.069, \:  \: \Delta S=0.14,\phantom{0}\: & \Delta C=0.14,\label{eq:scen-I6}\\
100~\text{fb}^{-1}: &
\Delta H= 0.052, \: \: \Delta S=0.035,\:  & \Delta C=0.035.\label{eq:scen-I100}
\end{eqnarray}
In principle these statistical errors could depend on the input value for $a$, $\theta$ and $\gamma$, but we will assume them constant for the remainder of this paper.

For a more detailed discussion of the experimental feasibility study, the reader is referred to Ref. \cite{Kristof}.

\boldmath
\section{Extraction of $\gamma$ at LHCb}\label{sec:gam}
\unboldmath
%
%
First information about $\gamma$ could in principle be obtained from $H$ as the 
general structure of this observable in (\ref{H-expr}) implies the following 
inequality \cite{FMR}:
\begin{equation}
H\geq\left[1-2\epsilon\cos^2\gamma +{\cal O}(\epsilon^2) \right]\sin^2\gamma
\end{equation} 
However, as the experimental data indicate a value of $H>1$, the corresponding bound
on $\gamma$ is not effective. 
 
\begin{figure}[htbp] 
\center
\includegraphics[width=3in]{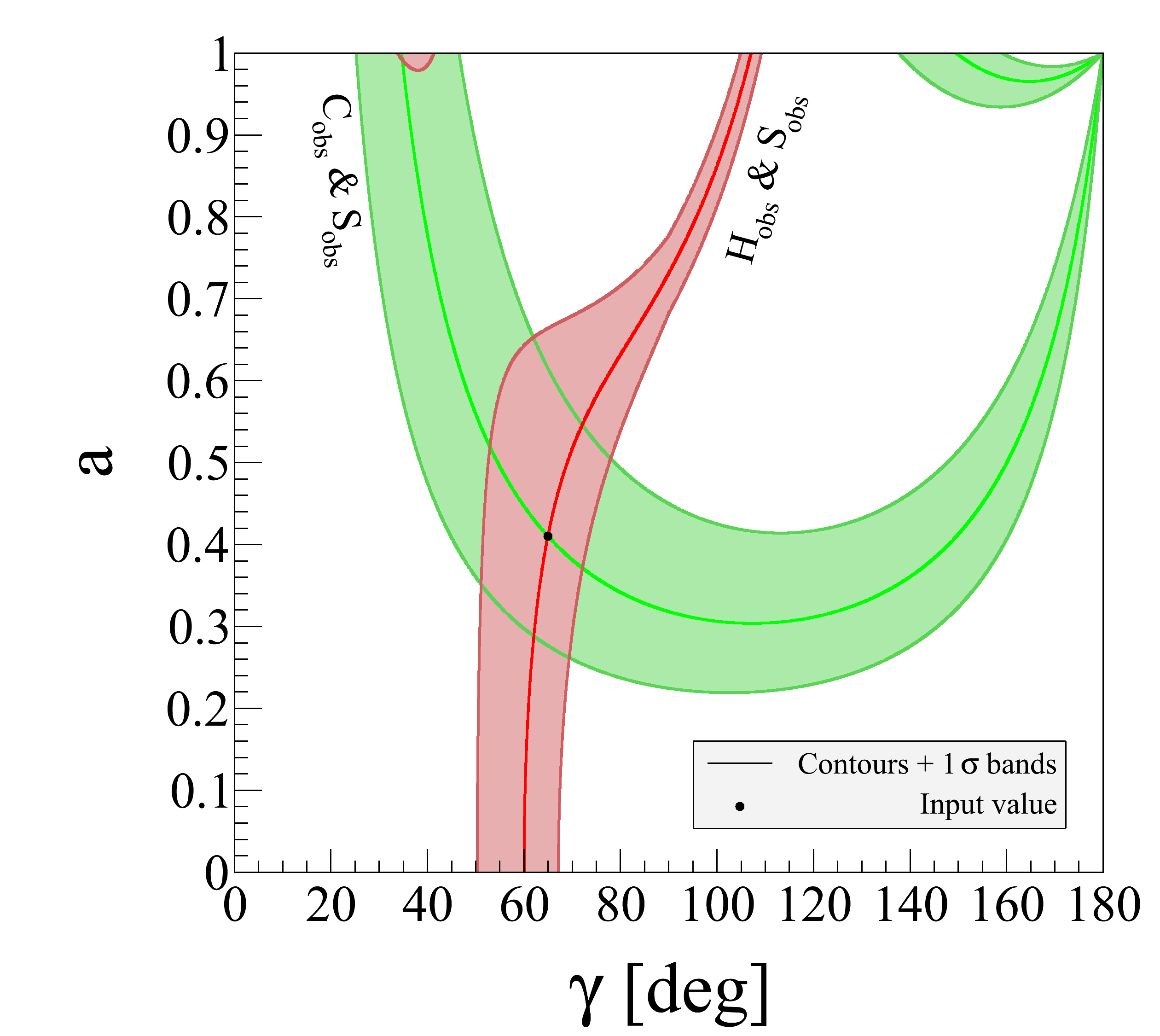} 
\includegraphics[width=3in]{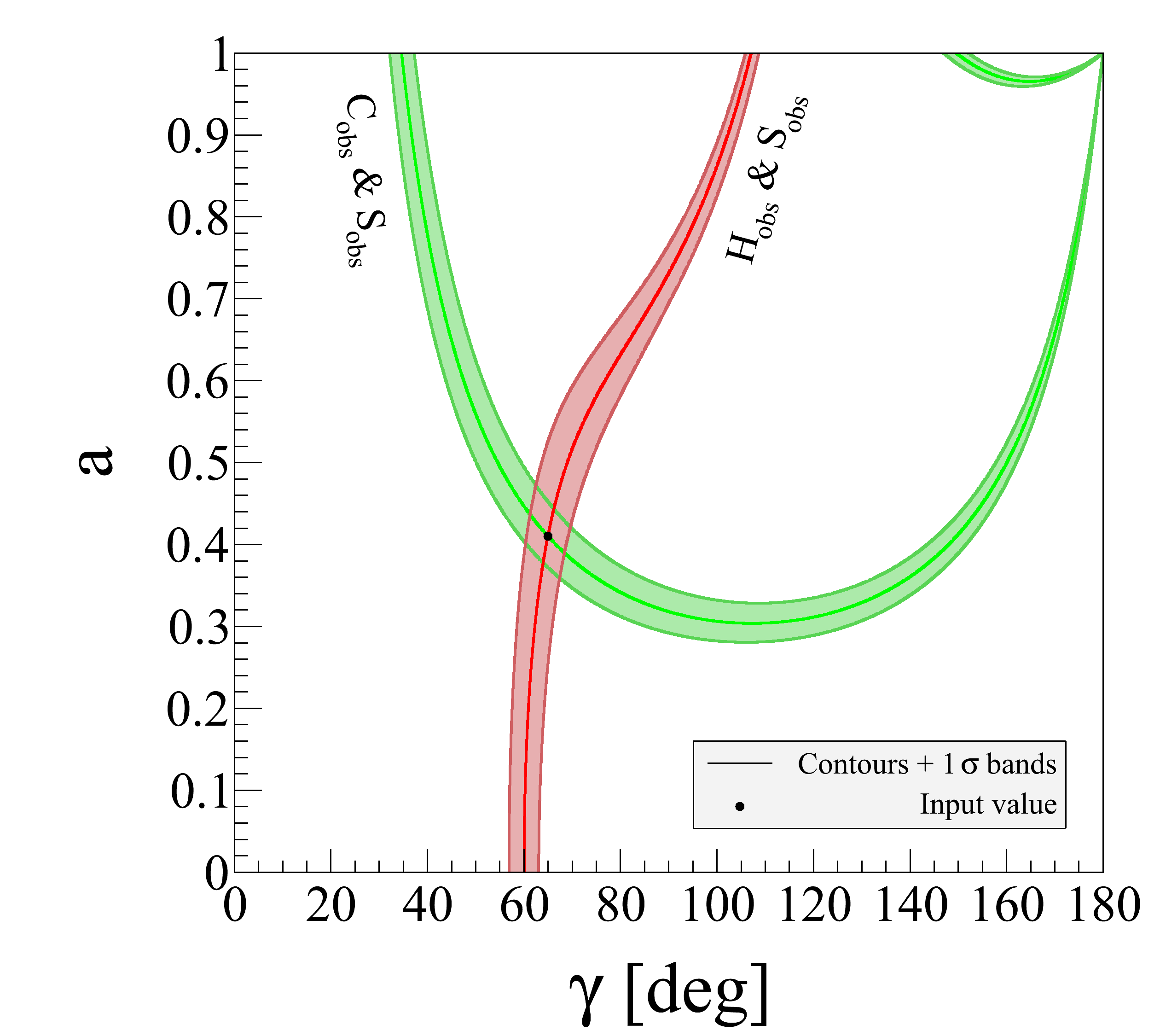} 
\caption{Determination of $\gamma$ and $a$ from intersecting contours. Situation resulting from
our LHCb feasibility study for 6 fb$^{-1}$ (left) and 100 fb$^{-1}$ (right).}
\label{fig:contours_gamma}
\end{figure}

As discussed in Ref.~\cite{RF-BspsiK}, using the $B^0_s$--$\bar B^0_s$ mixing phase
as an input, we can convert the $B_s^0\to J/\psi K_{\rm S}$ CP asymmetries $C$ and $S$ 
into a contour in the $\gamma$--$a$ plane, which is theoretically clean. Using the $U$-spin 
symmetry, we can determine another contour from $H$ and $S(B_s^0\to J/\psi K_{\rm S})$. 
Thanks to the $U$-spin relation in (\ref{rel-1}), the intersection of these curves allows us then 
to extract $\gamma$ and the penguin parameter $a$. Knowing these parameters, the 
CP-conserving strong phase $\theta$ can be extracted as well. In Fig.~\ref{fig:contours_gamma}, 
we illustrate the corresponding situation in the $\gamma$--$a$ plane for our LHCb 
study.

The angle $\gamma$ can also be extracted, together with $a$ and $\theta$, from a 
simultaneous $\chi^2$ fit to $C$, $S$ and $H$ using the expressions in
(\ref{Adir-def}), (\ref{Amix-def}) and (\ref{H-expr}). The corresponding numerical results
read as follows:
\begin{equation}\label{det-1}
6\,\mbox{fb}^{-1}:\quad 
\gamma= (65 \pm 10)^\circ, \quad 
a= 0.410 \pm 0.079, \quad \theta=(194\pm17)^{\circ},
\end{equation}
\begin{equation}\label{det-2}
100\,\mbox{fb}^{-1}:\quad \gamma= (65.0 \pm 3.2)^\circ, \quad
a= 0.410 \pm 0.023, \quad \theta=(194.0\pm4.1)^{\circ}.
\end{equation}
Here we give only the statistical errors, which show that LHCb should have sensitivity
to extract $\gamma$ from the $U$-spin-related $B_{s,d}\to J/\psi K_{\rm S}$ decays.
Using other avenues to extract $\gamma$, in particular theoretically
clean strategies employing $B_s^0\rightarrow D_s^{\mp} K^{\pm}$ and similar modes, LHCb
expects the following accuracies for $\gamma$:
\begin{equation}\label{eq:gamma_predicted}
6~\text{fb}^{-1}:\quad 
\gamma= (65.0 \pm 3.1)^\circ~\mbox{\cite{Roadmap}}, \quad 100~\text{fb}^{-1}:\quad 
\gamma= (65.0 \pm 1.0)^\circ~\mbox{\cite{LHCbUpgrade}},
\end{equation}
which should be compared with the current situation emerging from $B\to D^{(*)}K^{(*)}$
analyses:
\begin{equation}\label{gamma-current}
\gamma=(71^{+21}_{-25})^\circ~\mbox{(CKMfitter \cite{CKMfitter})}, \quad
(78\pm12)^\circ~\mbox{(UTfit \cite{UTfit})}.
\end{equation}
Consequently, LHCb should make significant progress with the determination of $\gamma$
with respect to the unsatisfactory current situation.

Concerning the $B_{s,d}\to J/\psi K_{\rm S}$ strategy, we find a good sensitivity for this 
angle in our feasibility study. We have also to deal with theoretical uncertainties originating 
from $U$-spin-breaking effects. As far as the corrections to (\ref{rel-1}) are concerned, they
actually play a very minor role for the determination $\gamma$ as the $a'$ terms in
(\ref{H-expr}) are strongly suppressed by the tiny $\epsilon$ parameter in \cite{RF-BspsiK}; numerically, they give a correction for $\gamma$ well below $1^\circ$. However,  
$U$-spin-breaking effects enter also through the ratio $|{\cal A}'/{\cal A}|$, which affects 
the determination of $H$ as defined in (\ref{H-expr}). We find that an uncertainty of 10\% for 
the relevant form-factor ratio 
$F_{B^0_dK^0}(M^2_{J/\psi};1^-)/F_{B^0_s\bar K^0}(M^2_{J/\psi};1^-)$, as 
considered in Ref.~\cite{FJFM}, corresponds to an error of about 
$\Delta\gamma|_{\rm FF}=12^\circ$, which essentially matches the statistical error of $10^\circ$
for our $6\,\mbox{fb}^{-1}$ scenario in (\ref{det-1}). For an LHCb upgrade with $100\,\mbox{fb}^{-1}$,
a form-factor uncertainty of  2.5\% would be required in order to match (\ref{det-2}), which serves 
as the ultimate benchmark for future calculations of this quantity. We hope that already by  
the end of the second LHCb data taking period, i.e.\ 2014--15, we will have a better 
understanding of $SU(3)$ breaking through studies of various other $U$-spin- and 
$SU(3)$-related $B$-meson decays, as well as through improved theoretical studies,
including in particular also non-factorisable $U$-spin-breaking corrections to 
$|{\cal A}'/{\cal A}|$. The current data from the Tevatron do not indicate sizable 
non-factorisable $SU(3)$-breaking effects \cite{RF-BsKK}, which is also supported by (\ref{Xi-res}). 

An important aspect of the determination of $\gamma$ through the $B_{s,d}\to J/\psi K_{\rm S}$
strategy is that it may reveal a  NP contribution to the $B^0_s\to J/\psi K_{\rm S}$ decay 
amplitude. This would be indicated by a value of $\gamma$ in conflict with the clean 
determinations through the pure ``tree" decays summarised in (\ref{eq:gamma_predicted}). 
In the following discussion, we assume that we have negligible NP effects of this kind, i.e.\
assume the SM expressions for the decay amplitudes in (\ref{Bs-ampl}) and
(\ref{Bd-ampl2}). 

\boldmath
\section{Clean Determination of the $B_s^0\to J/\psi K_{\rm S}$ Penguin 
Parameters at LHCb}\label{sec:pen}
\begin{figure}[tbp] 
\center
\includegraphics[width=3in]{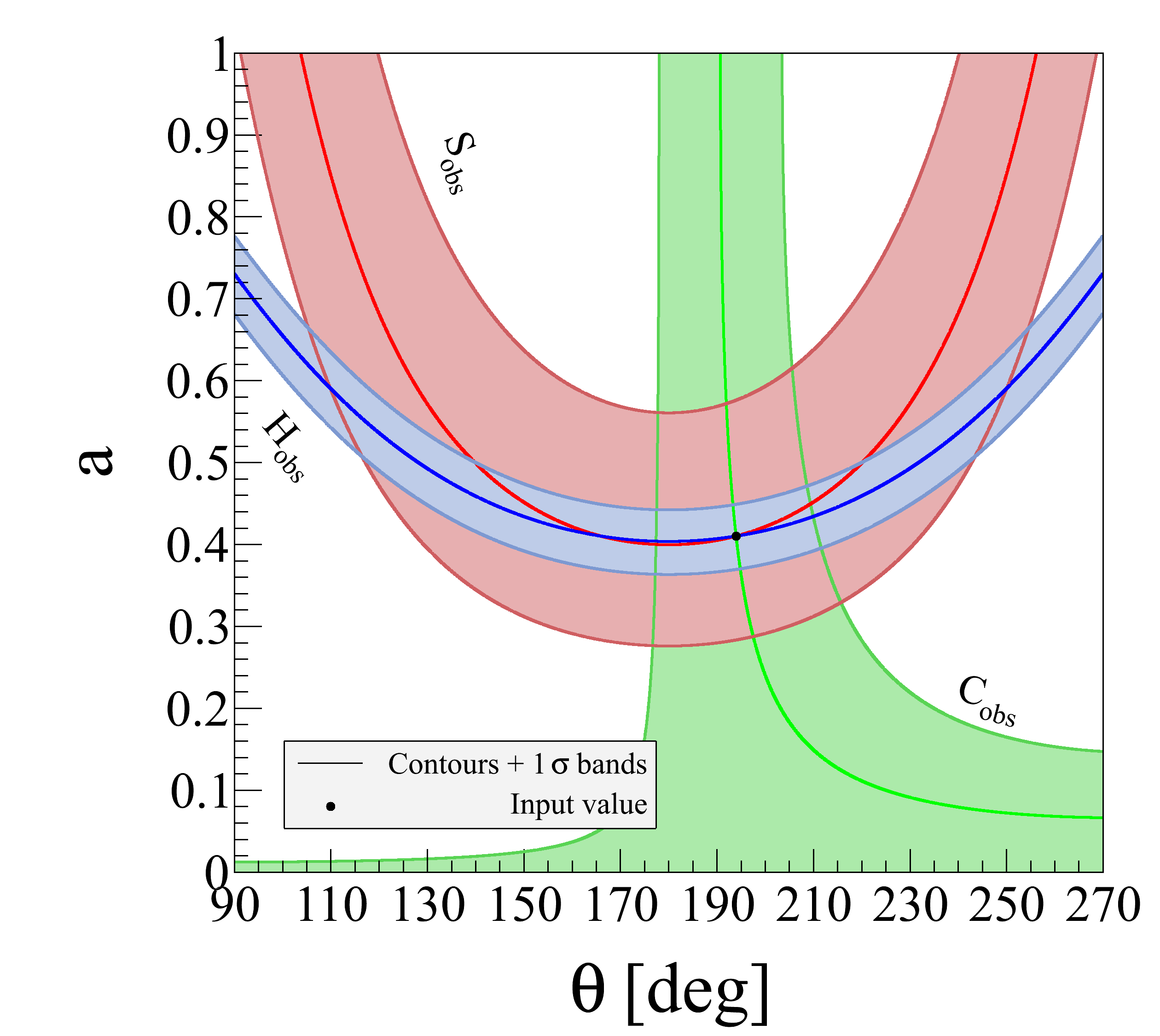} 
\includegraphics[width=3in]{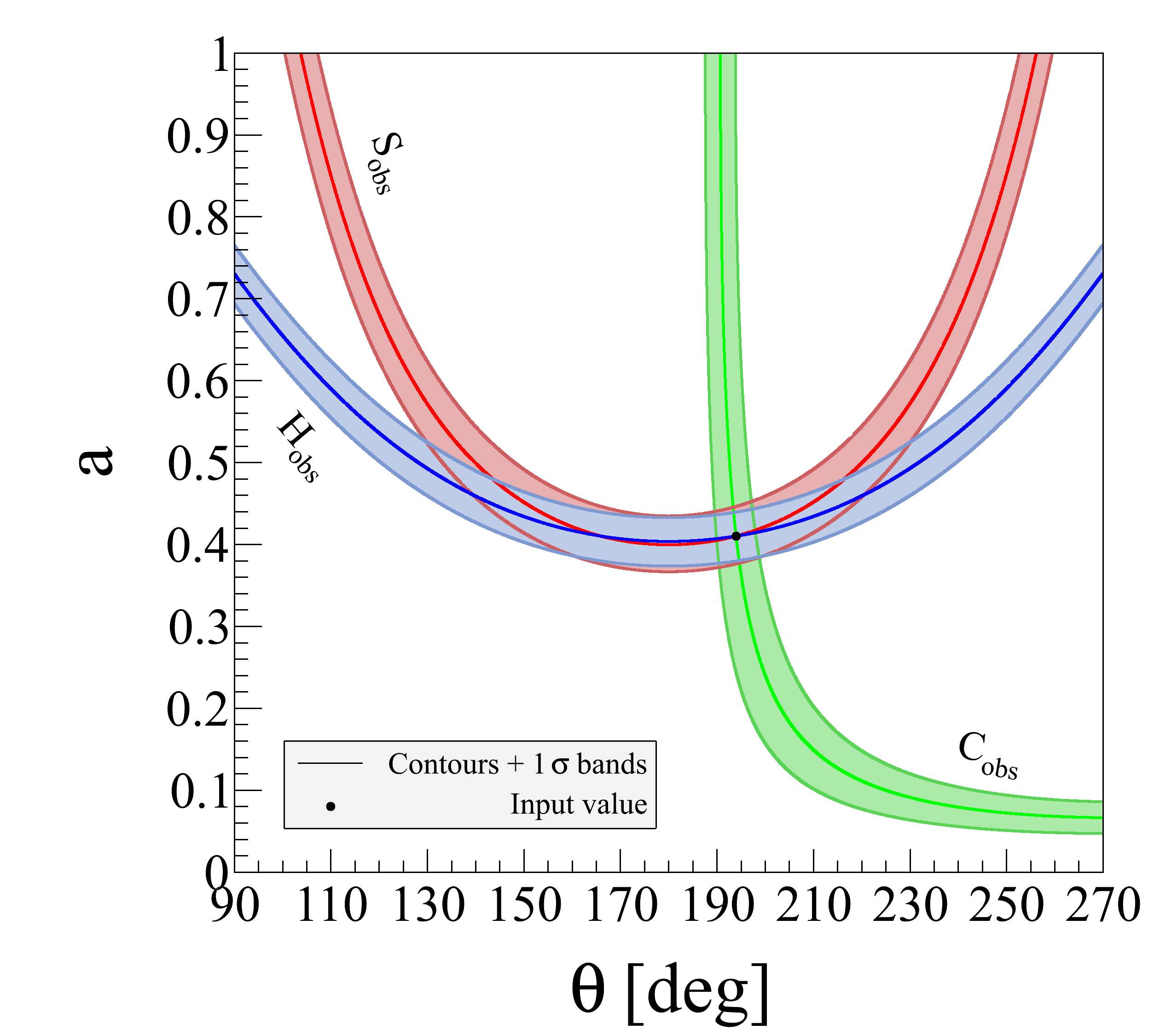} 
\caption{Determination of $a$ and $\theta$ from intersecting contours. Situation resulting from
our LHCb feasibility study for 6 fb$^{-1}$ (left) and 100 fb$^{-1}$ (right).}
\label{fig:contours_a_theta}
\end{figure}

\unboldmath
%
%
The major application of the $B_s^0\to J/\psi K_{\rm S}$ channel at LHCb will be the 
extraction of the hadronic penguin parameters $(a,\theta)$ and to take them into account
in the determination of the $B^0_d$--$\bar B^0_d$ mixing phase $\phi_d$ from 
$(\sin 2\beta)_{J/\psi K_{\rm S}}$. Since our goal is to minimise here the $U$-spin-breaking 
corrections, we shall refrain from using the $H$ observable and will use $\gamma$ as 
given in (\ref{eq:gamma_predicted}) as an input. Following Ref.~\cite{FJFM}, we can 
then convert the CP asymmetries $S$ and $C$ into theoretically clean contours in the 
$\theta$--$a$ plane; their intersection allows us then to extract $a$ and $\theta$. 
In Fig.~\ref{fig:contours_a_theta}, we illustrate the corresponding situation arising from 
our LHCb feasibility study. In order to guide the eye, we show also the contour corresponding 
to $H$, which would be interesting to obtain insights into $U$-spin-breaking 
effects. Performing again a $\chi^2$ fit yields the following results:
\begin{eqnarray}
6~\text{fb}^{-1}: & a= 0.41 \pm 0.14\,(\text{stat.})\pm 0.02\,(\gamma) \: , \: & 
\theta=\left[194 \pm 16\,(\text{stat.})\pm 0.8\,(\gamma)\right]^{\circ}, \\
100~\text{fb}^{-1}: & a= 0.41 \pm 0.03\,(\text{stat.})\pm 0.01\,(\gamma)\: , \: & 
\theta=\left[194.0\pm 4.0\,(\text{stat.})\pm 0.3\,(\gamma)\right]^{\circ}.
\end{eqnarray}
We observe that LHCb is expected to be able to determine the $B^0_s\to J/\psi K_{\rm S}$
penguin parameters with impressive precision. It should be emphasised that these parameters
rely only on the SM expression for the decay amplitude and are theoretically clean.

\begin{figure}[tbp] 
\centering
\includegraphics[width=3in]{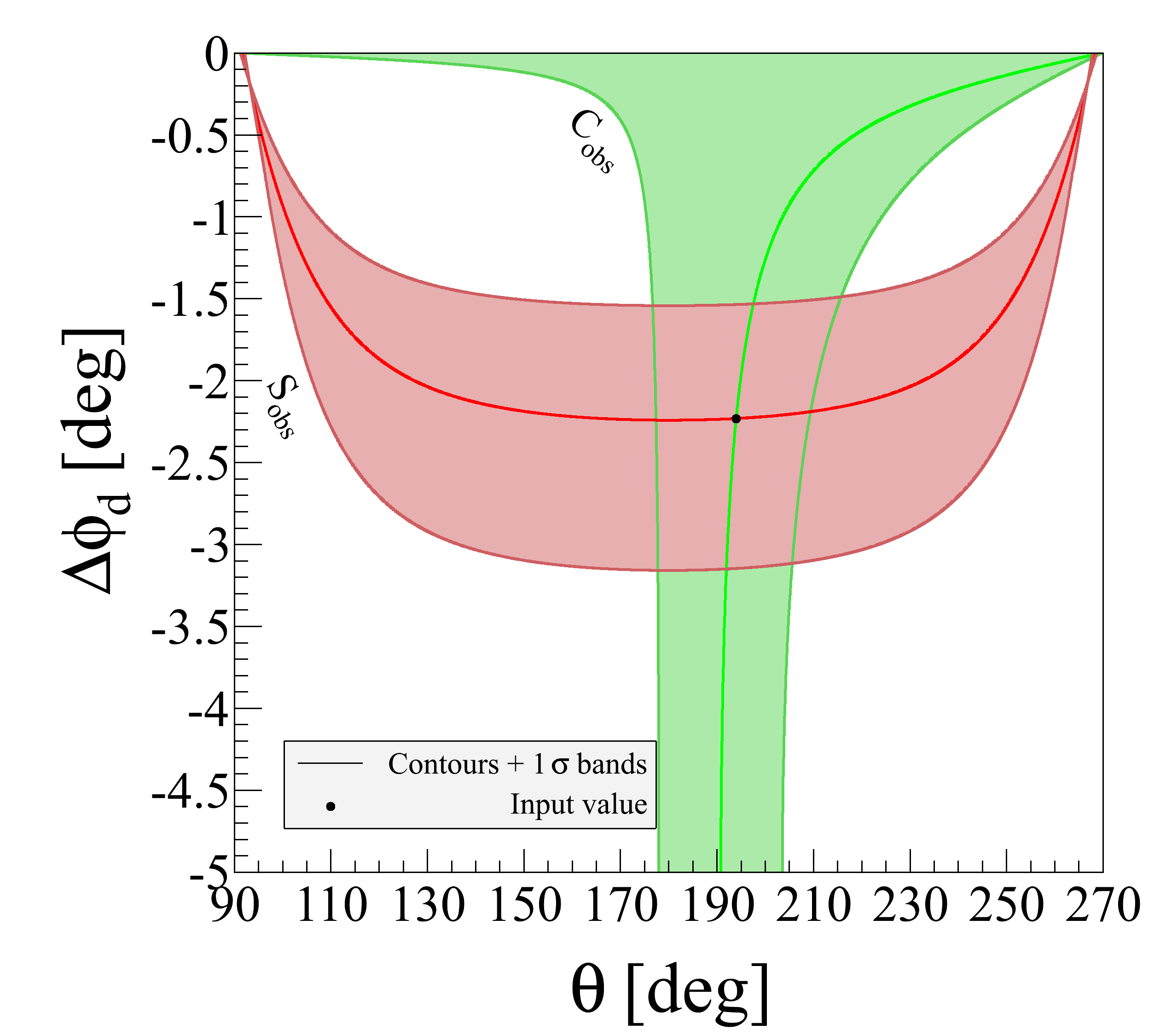}
\includegraphics[width=3in]{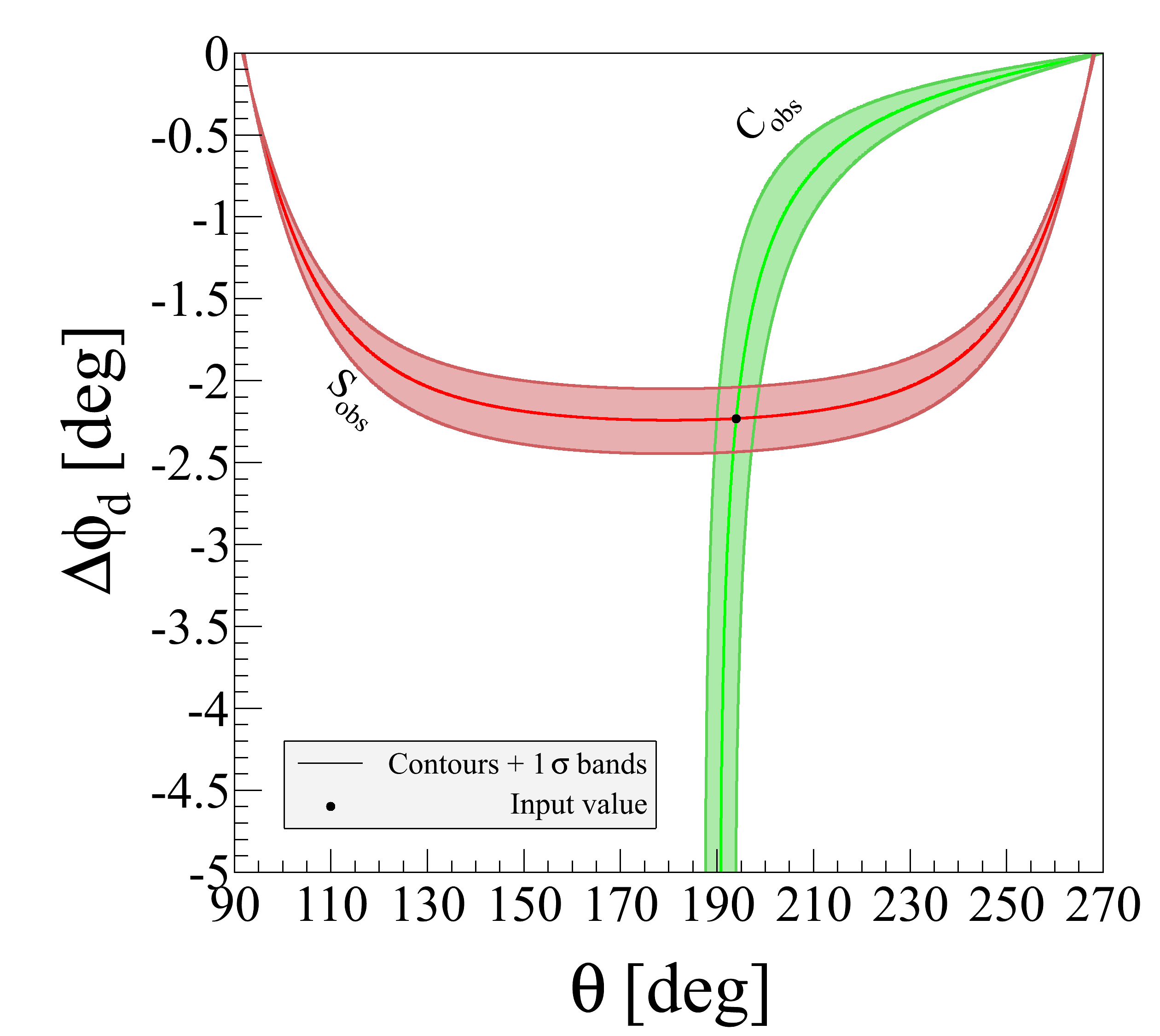} 
\caption{Determining the penguin correction $\Delta\phi_d$ to $\sin(2\beta)$ from intersecting contours. Situation for 6 fb$^{-1}$ (left) and 100 fb$^{-1}$ (right).}\label{fig:DelPhi}
\end{figure}

\boldmath
\section{Controlling Penguin Effects in the Measurement of 
$(\sin 2\beta)_{J/\psi K_{\rm S}}$ at LHCb}\label{sec:pen-cont}
\unboldmath
%
%
The values of the penguin parameters $a$ and $\theta$ are interesting from the point of
view of strong interaction dynamics. However, they have also an important practical application
for testing the Kobayashi--Maskawa mechanism of CP violation. 
Using the $U$-spin relation (\ref{rel-1}), we can relate the $B^0_s\to J/\psi K_{\rm S}$ parameters
to their $B^0_d\to J/\psi K_{\rm S}$ counterparts, which allows us then to control 
the hadronic uncertainty of the measurement of the $B_d^0$--$\bar B_d^0$ mixing 
phase $\phi_d$ through the mixing-induced CP violation in $B_d^0\to J/\psi K_{\rm S}$. 
The corresponding, generalised expression reads as follows \cite{FJFM}:
\begin{equation}\label{S-gen}
\frac{S(B_d^0\to J/\psi K_{\rm S})}{\sqrt{1-C(B_d^0\to J/\psi K_{\rm S})^2}}
=\sin(\phi_d+\Delta\phi_d),
\end{equation}
where
 \begin{eqnarray}
\sin\Delta\phi_d&=&\frac{2 \epsilon a\cos\theta \sin\gamma+\epsilon^2a^2
\sin2\gamma}{N\sqrt{1-C(J/\psi K_{\rm S})^2}}\label{sDelPhi}\\
\cos\Delta\phi_d&=&\frac{1+ 2 \epsilon a\cos\theta \cos\gamma+\epsilon^2a^2
\cos2\gamma}{N\sqrt{1-C(J/\psi K_{\rm S})^2}}\label{cDelPhi}
\end{eqnarray}
with $N\equiv1+2\epsilon a \cos\theta \cos\gamma+\epsilon^2a^2$,
so that
\begin{equation}\label{tDelPhi}
\tan\Delta\phi_d=\frac{2 \epsilon a\cos\theta\sin\gamma+\epsilon^2a^2
\sin2\gamma}{1+ 2 \epsilon a\cos\theta\cos\gamma+\epsilon^2a^2\cos2\gamma}.
\end{equation}
The current value of the mixing-induced CP violation is given by
\begin{equation}\label{SpsiK-curr}
S(B_d^0\to J/\psi K^0)=0.655 \pm 0.024,
\end{equation}
while the direct CP asymmetry is currently measured as
\begin{equation}\label{C-psiK}
C(B_d^0\to J/\psi K^0)=-0.003 \pm 0.020. 
\end{equation} 
These numbers are averages of the BaBar and Belle data over the 
$B_d^0\to J/\psi K_{\rm S}$ and $B_d^0\to J/\psi K_{\rm L}$ final states, as obtained by the
Heavy Flavour Averaging Group \cite{HFAG}. The value in (\ref{C-psiK}) implies
that the deviation of the square root in (\ref{S-gen}) from one is at most 
at the level of $0.0002$, i.e.\ completely negligible.

\begin{figure}[tbp] 
\centering
\includegraphics[width=3in]{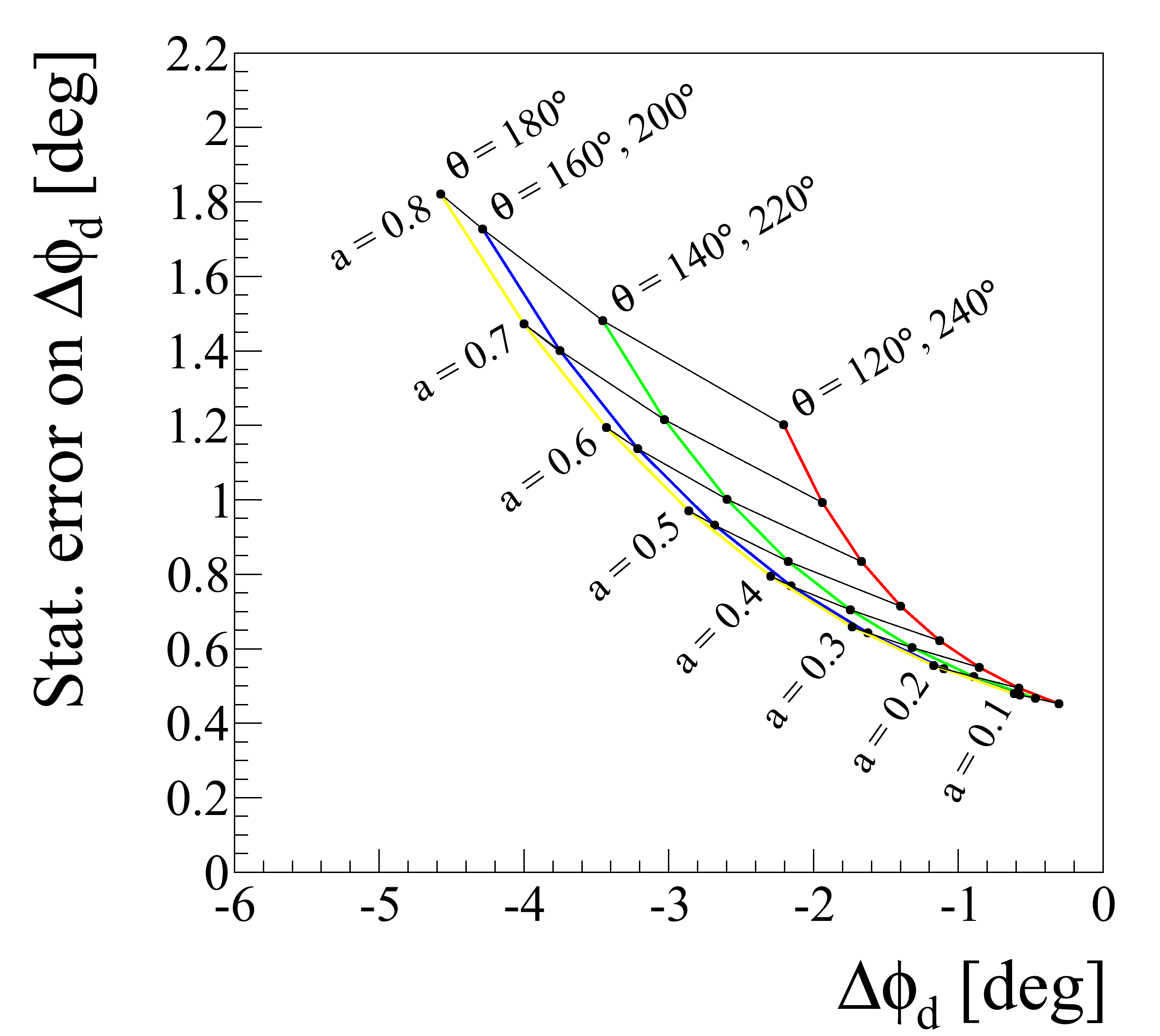}
\includegraphics[width=3in]{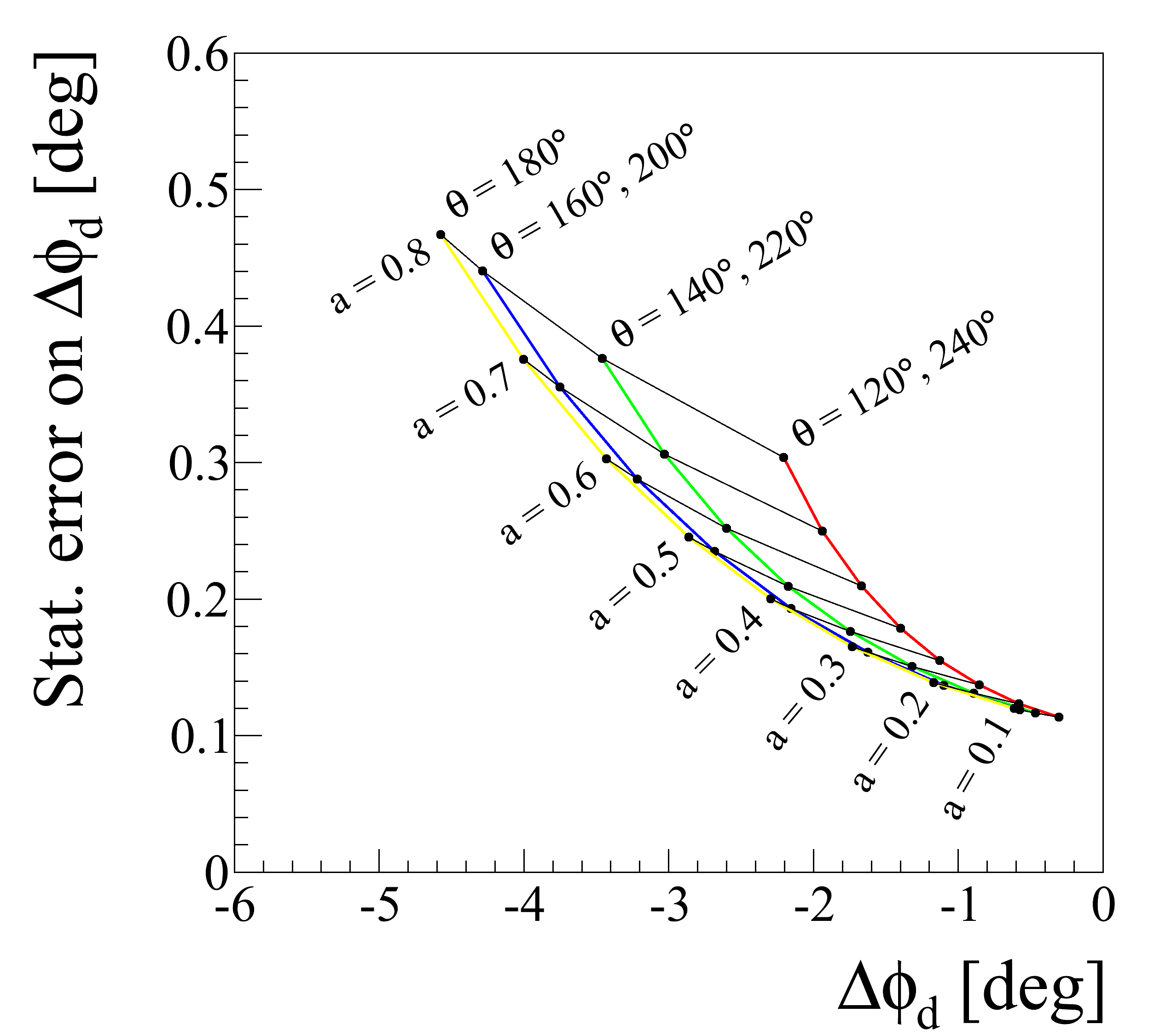} 
\caption{Correlation between the hadronic shift $\Delta\phi_d$ and its error for our
LHCb feasibility study for different input values of the penguin parameters $a$ and
$\theta$. Situation for 6 fb$^{-1}$ (left) and 100 fb$^{-1}$ (right); note the different
scales of the $y$ axes.}\label{fig:DelPhi-corr}
\end{figure}

Using these formulae, we can convert the contour plots in Fig.~\ref{fig:contours_a_theta}
into curves in the $\theta$--$\Delta\phi_d$ plane, as shown in Fig.~\ref{fig:DelPhi}. In this
analysis, we have uncertainties from $U$-spin-breaking corrections to (\ref{rel-1}), 
which we parametrise as follows:
\begin{equation}\label{U-break}
a = \xi a' \quad,\quad \theta = \theta'+\Delta\theta.
\end{equation}
If we vary $\xi\in[0.85,1.15]$ and 
$\Delta\theta\in[-20^{\circ},20^{\circ}]$, and perform a $\chi^2$ fit to 
$C$ and $S$, we obtain
\begin{equation}\label{pen-1}
~6~\text{fb}^{-1}: \Delta\phi_d=\left[-2.23\pm 0.78~(\text{stat.})\pm 0.12~(\gamma) ~^{+0.29}_{-0.40} (\xi) ~^{+0.33}_{-0.07} (\Delta\theta) \right]^{\circ}, 
\end{equation}
\begin{equation}\label{pen-2}
100~\text{fb}^{-1}:  \Delta\phi_d=\left[-2.23\pm 0.19~(\text{stat.})\pm 0.04~(\gamma) ~^{+0.29}_{-0.40} (\xi) ~^{+0.33}_{-0.07} (\Delta\theta) \right]^{\circ}.
\end{equation}
In order to study the dependence on the input values of the penguin parameters for
our analysis, we show the correlation between the hadronic shift $\Delta\phi_d$ and
its statistical error at LHCb in Fig.~\ref{fig:DelPhi-corr}. The corresponding curves show
nicely that we can actually well determine the $\Delta\phi_d$ correction for a wide range of
$(a,\theta)$ that should contain the ``true" values of these parameters. It should also be noted
that already a small penguin contribution with $a=0.1$ gives a correction of 
$\Delta\phi_d\sim -0.5^\circ$.

In order to fully appreciate the results of this study, we have to look at the precision for 
$(\sin2\beta)_{J/\psi K_{\rm S}}$ at LHCb: we expect $0.022$ for $2\:\rm fb^{-1}$~\cite{LHCb}, 
and  $0.003$--$0.010$ (depending on to be determined systematic errors) for 
$100\:\rm fb^{-1}$~\cite{LHCbUpgrade}. We extrapolate these numbers into 
$0.014$ for $6\:\rm fb^{-1}$, 
which should be compared with (\ref{SpsiK-curr}). For a central value of 
$S(B_d^0\to J/\psi K^0)\sim 0.655$, these numbers correspond to errors of about 
$1^\circ$ and $(0.2\mbox{--}0.8)^\circ$ for the phase itself. 
In order to match these very impressive precisions, 
we definitely have to control the doubly Cabibbo-suppressed penguin contributions in 
$B^0_d\to J/\psi K_{\rm S}$. Looking at the numerical results in (\ref{pen-1}) and (\ref{pen-2}), 
we observe that we can actually achieve this goal. Already with $6\:\rm fb^{-1}$ we have
to control the penguin effects, for an upgrade of LHCb this is absolutely necessary to 
fully exploit the tremendous statistics. Measurements along these lines may eventually
allow us to resolve CP-violating NP contributions to $B^0_d$--$\bar B^0_d$ mixing. 

Let us finally note that we can -- in addition to the direct CP violation in the 
\mbox{$B_d^0\to J/\psi K_{\rm S,L}$} channels \cite{FJFM} -- add another observable 
to this analysis. It is the direct CP asymmetry of the 
$B_u^+\to J/\psi \pi^+$ decay, which we get from the $B_s^0\to J/\psi K_{\rm S}$ topologies 
in Fig.~\ref{fig:BstoJpsiKS} by replacing the strange spectator quark by an up quark. 
The resulting contours in 
Fig.~\ref{fig:DelPhi} are similar to those related to $C$. Already with $1\:\rm fb^{-1}$, as 
expected by the end of 2011, we find from a feasibility study similar to that described above 
that the $B_u^+\to J/\psi \pi^+$ mode should be clearly visible at
LHCb, with a statistical error of $C(B_u^\pm\to J/\psi \pi^\pm)=0.005$. 
However, the production asymmetry between $B$ and $\bar B$ states at the LHC 
(which, unlike the Tevatron, is a $pp$ collider) is expected to be non-negligible and 
will have to be measured and corrected for. This is likely to induce irreducible systematic 
errors at a few per-cent level for the measurement of $C(B_u^\pm\to J/\psi \pi^\pm)$, which 
have to be studied in further detail.

\section{Conclusions}\label{sec:concl}
%
%
As was pointed out about one decade ago, the decay $B^0_s\to J/\psi K_{\rm S}$ is the 
$U$-spin partner of the ``golden" mode $B^0_d\to J/\psi K_{\rm S}$ and allows an 
extraction of the UT angle $\gamma$ and of the corresponding penguin parameters.  
In the summer of 2010, the CDF collaboration has 
announced the first observation of the $B^0_s\to J/\psi K_{\rm S}$ transition, with a 
branching ratio in agreement with an $SU(3)$ relation to the $B_d^0\to J/\psi \pi^0$ 
decay. 

We have calculated a correlation between the mixing-induced CP violation  in the
$B^0_s\to J/\psi K_{\rm S}$ channel, $S$, and $\sin\phi_s$, which serves as a ``target region" for 
the first CP violation measurement in this decay. We observe an interesting pattern
for values of the $B^0_s$--$\bar B^0_s$ mixing phase $\phi_s$ in the region of the
current measurements from the Tevatron: we can distinguish between the twofold solutions
for $\phi_s$ arising from the measurement of $\sin\phi_s$, with $S\sim0$ and $S\sim-0.8$
for $\phi_s\sim-30^\circ$ and $\phi_s\sim-150^\circ$, respectively, whereas we expect
$S\sim+0.5$ in the SM. Thanks to improved measurements of input parameters,
these predictions can be sharpened in the future.

By the end of 2011, i.e.\ the end of the first LHCb data taking period, we expect a clear signal
for $B^0_s\to J/\psi K_{\rm S}$, with very crude first information on the corresponding
CP asymmetries. Extrapolating from published LHCb studies, we have performed a detailed
feasibility study for the measurement of the $B^0_s\to J/\psi K_{\rm S}$ observables, 
yielding about 4000 and 70000 signal events for $6\:\rm fb^{-1}$ and $100\:\rm fb^{-1}$, 
respectively. These integrated luminosities refer to the end of the second LHCb data 
taking period of 2014--15 and a subsequent LHCb upgrade. Using input parameters as 
suggested by current data, we have shown that the $B_{s,d}\to J/\psi K_{\rm S}$ strategy 
offers indeed another determination of $\gamma$ for LHCb. However, we conclude that the most 
important application of this method will be the determination of the corresponding 
hadronic penguin parameters and their control in the measurement of 
$(\sin2\beta)_{J/\psi K_{\rm S}}$, which will allow us to match the corresponding 
experimental precision. Studies along these lines may eventually 
allow us to resolve NP in $B^0_d$--$\bar B^0_d$ mixing.

\vspace*{0.5truecm}

\noindent
{\it Acknowledgements}\\
We would like to thank Marcel Merk and Wouter Hulsbergen for carefully reading the manuscript
and useful comments and suggestions. 


\end{document}